\documentclass[journal=jctcce,manuscript=article]{achemso}
\usepackage{amssymb,amsmath,amsfonts,multicol,multirow,longtable,array,mathpazo}
\usepackage{lscape}
\usepackage[version=3]{mhchem} 
\usepackage{appendix}
\usepackage{soul} 
\usepackage[usenames]{color}
\usepackage{makecell}
\usepackage{enumerate}

\usepackage{xr}

\usepackage[T1]{fontenc} 

\usepackage{booktabs}
\usepackage{tikz}

\usepackage{threeparttable}
\usepackage{graphicx}
\usepackage{subfigure}

\usepackage{colortbl}
\usepackage{framed}
\usepackage{verbatim}
\usepackage{amsbsy}
\usepackage{bm}
\usepackage{bbm}
\usepackage[normalem]{ulem}
\usepackage{float}
\usepackage[linesnumbered,boxed]{algorithm2e}
\usepackage{algorithm2e}

\newcounter{xscheme}

\newfloat{Algorithm}{htbp}{alg}
\floatname{Algorithm}{Algorithm}

\newcounter{exe}[figure]
\newcommand{\iexe}{\refstepcounter{exe}\the\value{exe}:}

\setkeys{acs}{maxauthors = 0} 

\author{Zikuan Wang}
\affiliation{Qingdao Institute for Theoretical and Computational Sciences, Institute of Frontier and Interdisciplinary Science, Shandong University, Qingdao, Shandong 266237, P. R. China}
\author{Wenjian Liu}\email{liuwj@sdu.edu.cn}
\affiliation{Qingdao Institute for Theoretical and Computational Sciences, Institute of Frontier and Interdisciplinary Science, Shandong University, Qingdao, Shandong 266237, P. R. China}

\title{iOI: an Iterative Orbital Interaction Approach for Solving the Self-Consistent Field Problem}

\setkeys{acs}{articletitle=true}

\begin{document}

\begin{abstract}
An iterative orbital interaction (iOI) approach is proposed to solve, in a bottom-up fashion,
the self-consistent field problem in quantum chemistry.
While it belongs grossly to the family of fragment-based quantum chemical methods, iOI is distinctive in that
(1) it divides and conquers not only the energy but also the wave function, and that
(2) the subsystems sizes are automatically determined by successively merging neighboring small subsystems until they are just enough for converging the wave function to a given accuracy.
Orthonormal occupied and virtual localized molecular orbitals are obtained in a natural manner, which can be used for all post-SCF purposes.
\end{abstract}

keywords: iterative orbital interaction, self-consistent field, bottom-up solver, fragment localized molecular orbitals, modified projected hybrid orbitals,
automatic fragmentation and merging, iterative matrix block-diagonalization

\maketitle

\section{Introduction}
The molecular orbitals (MO) of a chemical system resulting from a self-consistent field (SCF) calculation can always be localized as long as there exists a gap between the occupied and virtual spaces.\cite{ACR-FLMO,LMOChemRev}
Because of this, the electronic structure of a local part of a gapped large molecule (macromolecule or molecular cluster) is not much affected by those parts that are far away in distance.
Such locality (also called ``nearsightedness''\cite{Nearsightedness}) is fundamental to all low-scaling quantum chemical methods.
The issue is how to materialize such locality in a bottom-up fashion (i.e., without invoking the expensive top-down localization\cite{ACR-FLMO,LMOChemRev} of the global occupied and virtual canonical MOs).
As a matter of fact, the best answer to this was already provided by chemists in the early days of chemistry: as the basic building units,
functional groups or fragments in general reflect best the locality of chemical systems. As such,
various fragment-based quantum chemical methods have been developed in the last decades, including
mixed quantum mechanics/molecular mechanics (QM/MM)\cite{QMMM1,EFP1996,EFP2001,X-Pol2007},
multilayer QM/QM\cite{ONIOM1,ONIOM2,XO2010,X-Pol2012}, divide-and-conquer (DC)\cite{YangDC1991,YangDC1995,NakaiDC2007,MerzDC2010},
embedding\cite{SubDFT1993,LSCF1996,SubDFT2006,FDE-pot2010,FDE-EO2015,Carter2011quantum,MFEmbedding2012,DMET2012,DMET2013,HoffmannEmbedding2014,Bootstrap2016,WFTinDFT-LMO,OCBSE2017,o-efsDFT2019,ChemPotEmb2019,QM-ELMO2021}, and
orbital-\cite{SCF-MI1980,Incremental1,Incremental2,CIM2002,CIM2010,NLCC2004,NLCC2008,DEC2010,XPol-SAPT2011,ElongationPCCP2012} and energy-based\cite{CG-MTA1994,CG-MTA2006,FMO1999,huang2005kernel,MFCC2003,MetalMFCC2019,SMFA2005,SMFA2012,
jiang2006electrostatic,GEBF1,GEBF2,EE-MBE2007a,EE-MBE2007b,MFBA2010,HMBI2010,MIM2011,MOBE2012,GenMBE2012,le2012combined,DOD2014} fragmentation schemes,
which differ mainly in how to divide the entire system into fragments and how to conquer their mutual interactions, so as to achieve
high accuracy yet with low cost. For instance, in the energy-based fragmentation approaches,
the electronic energy and energy-related properties of the entire system are obtained simply by assembling the fragmental values in one way or another\cite{FragChemRev2012,MFCC-ACR2014,GEBF-ACR2014,SMFA-ACR2014,FMOPCCP2014,gadre2014quantum,FragChemRev2015,raghavachari2015accurate,gordon2017fragmentation}.
The accuracy can further be improved greatly by invoking
a low-level treatment of the entire system\cite{MIMMOB2015} or by using a superposition of multiple fragmentation topologies\cite{LC-MFT2019}.
Of course, instead of making direct use of such chemical intuitions, one can also exploit locality by purely mathematical techniques such as prescreening\cite{NAOlinearscaling,linearscaling1999,linearscaling2012,linearscaling2013}.

It is not surprising that the above fragment-based methods can also be used to describe low-lying excited states that are localized
in a small region of the system\cite{FMO-TDDFT,HoffmannEmbeddingExcited2010,SS-EP2013,PE-MCSCF2013,DCexcited2013,WFTinDFT2014,subTDDFT2015,EFPexcited2016,GEBFexcited,EE-GMF2019,MLMBE,ML-MLMBE,QM-ELMO2020}.
Low-lying delocalized excited states (excitons or transitions between delocalized orbitals)
can be accessed either by linear combinations of local states\cite{REM-PRB2005,REM-JCC2012,REM-JCP2012,ALMO-TDDFT2015,ALMO-TDDFT2016,ALMO-CIS2015,FDE-EO-TDDFT2016}
or by enlarged buffer regions\cite{DCexcited2017}.
Short-range charge transfer (CT) effects can also be accounted for deliberately to improve those states dominated by local excitations\cite{ALMO-CIS+CT2017}.
Given such successes, it must be admitted that such descriptions of excited states all rely on some information known in advance.
It is often the case that the so-calculated excitation energies are sufficiently good but transition properties that are sensitive to the tails of the wave functions
are still problematic. Moreover, genuine CT excitations are missed by construction by fragment-based methods, needless to say high-lying states, the radial extensions of which are usually far beyond the intuitively partitioned fragments.
For such types of states, one should generate converged localized molecular orbitals (LMO) in a bottom-up fashion (to stay within the fragmentation picture)\cite{ACR-FLMO}
and make proper use of both the energetic and spatial localities of particle-hole pairs (the real basis of post-SCF approaches)
for describing excited states of a given energy range\cite{Triad}.

Stimulated by the iterative configuration interaction (iCI) approach\cite{SDS,iCI,iCIPT2,iCIPT2New}
for solving the many-electron Schr\"odinger equation,
we propose here an iterative orbital interaction (iOI) approach for solving the
Hartree-Fock/Kohn-Sham (HF/KS) equation in a bottom-up fashion. Since the HF/KS equation represented in a finite basis
can, at each iteration, be viewed as
a one-electron full CI problem, the same strategy employed by iCI, i.e., projecting the Hamiltonian onto a small set of
physically oriented basis functions (linear combinations of determinants or configuration state functions, the number of which
grows along macroiterations), can also be used in iOI,
in which the ``physically oriented basis functions'' are just linear combinations of primitive fragment LMOs (pFLMO) from subsystem calculations\cite{FLMO1,FLMO3}.
By merging successively neighboring small fragments into larger ones along the macroiterations, the pFLMOs are improved steadily and many of them
are essentially converged.
They provide good initial guess for the final global SCF calculation such that the number of SCF iterations is reduced greatly. In short,
iOI belongs grossly to the family of fragment-based methods but is distinctive in that it
divides and conquers both the energy and wave function (orbitals). Orthonormal occupied and virtual LMOs
are obtained in a natural manner\cite{FLMO1,FLMO3}, which can be used for all post-SCF purposes.
The algorithm will be detailed in Sec. \ref{iOI}. Some showcases are then provided in Sec. \ref{Results}
to reveal the efficacy of iOI.


\section{iOI}\label{iOI}
The aim here is to determine the converged, orthonormal LMOs $\{\psi_p\}$ of a gapped molecule in a bottom-up fashion.
As usual, they are expanded in terms of atom-centered Gaussian basis functions $\{\chi_{\mu}\}$,
\begin{equation}
\psi_p(\boldsymbol{r}) =\sum_{\mu}^{M_p}\chi_{\mu}(\boldsymbol{r})C_{\mu p}[V_{eff}].\label{LMOexpan}
\end{equation}
The questions are how to identify \textit{a priori} those atomic orbitals (AO) $\{\chi_{\mu}\}$ that are most relevant for an LMO
$\psi_p(\boldsymbol{r})$ centered at $\boldsymbol{G}_p=\langle\psi_p|\boldsymbol{r}|\psi_p\rangle$ with a radius of $R_p$,
and how to determine the corresponding expansion coefficients $\mathbf{C}_p$ that depend on the effective potential $V_{eff}(\boldsymbol{r})$
under the constraints of the Brillouin conditions.
Since every basis function $\chi_{\mu}$ centered at $\boldsymbol{G}_{\mu}$ also has a finite radius $R_{\mu}$ of distribution, it is clear that
only those functions with distances $|\boldsymbol{G}_{\mu}-\boldsymbol{G}_p|$ smaller than $R_\mu+R_p$ contribute to the expansion \eqref{LMOexpan}.
However, such information cannot be employed to truncate \textit{a priori} the expansion \eqref{LMOexpan} since neither the center $\boldsymbol{G}_p$ nor the radius $R_p$
of $\psi_p$ is known in advance.
Nonetheless, a good guess on the center $\boldsymbol{G}_p$ of $\psi_p$ stems naturally from the concept of functional group or fragment in general.
Therefore, the issue becomes how to merge the initially partitioned fragments in a way that is just enough to determine the right basis functions
for expanding the set of LMOs localized on a given fragment.
As for the determination of the expansion coefficients $\mathbf{C}_p$, the short-range exchange-correlation part of the
effective potential $V_{eff}(\boldsymbol{r})$ is taken into account automatically during the merging process, whereas
the long-range electrostatic interaction from the environment can be accounted for explicitly
by means of, e.g., the multipolar expansion technique\cite{BDF1,BDFrev2020}. Nevertheless, when all LMOs of the system are to be targeted,
it is more convenient and accurate to account for the long-range interactions by a few global SCF iterations.

The above analysis suggests the following algorithm for the generation of the LMOs of the whole system:
\begin{enumerate}[(1)]
\item\label{group} Partition the whole system into nonoverlapping functional groups (e.g. \ce{CH3}, \ce{CH2}, \ce{C=O}, \ce{SH}, \ce{C-Cl}, etc.), each of which is obtained
by cutting the single bonds between main group elements other than hydrogen, halogens, and rare gases;
\item\label{fragment} Initialize the macroiteration count $\tilde{m}$ to 0. Join the functional groups defined in Step (\ref{group}) to
form $N_{\mathcal{F}}^{(\tilde{m})}$ nonoverlapping fragments $\mathcal{F}_L^{(\tilde{m})}$ ($L\in [1,N_{\mathcal{F}}^{(\tilde{m})}]$) according to chemical intuition, so that every atom in the whole
system belongs to one and only one fragment (see Fig.\ref{ala4} for an illustration). The nonoverlapping fragments are called here primitive
fragments (pFrag), which form precisely the whole system and are typically of 10 to 30 atoms for the best cost-performance ratio;
\begin{figure}
\centering
\includegraphics[width=0.7\textwidth]{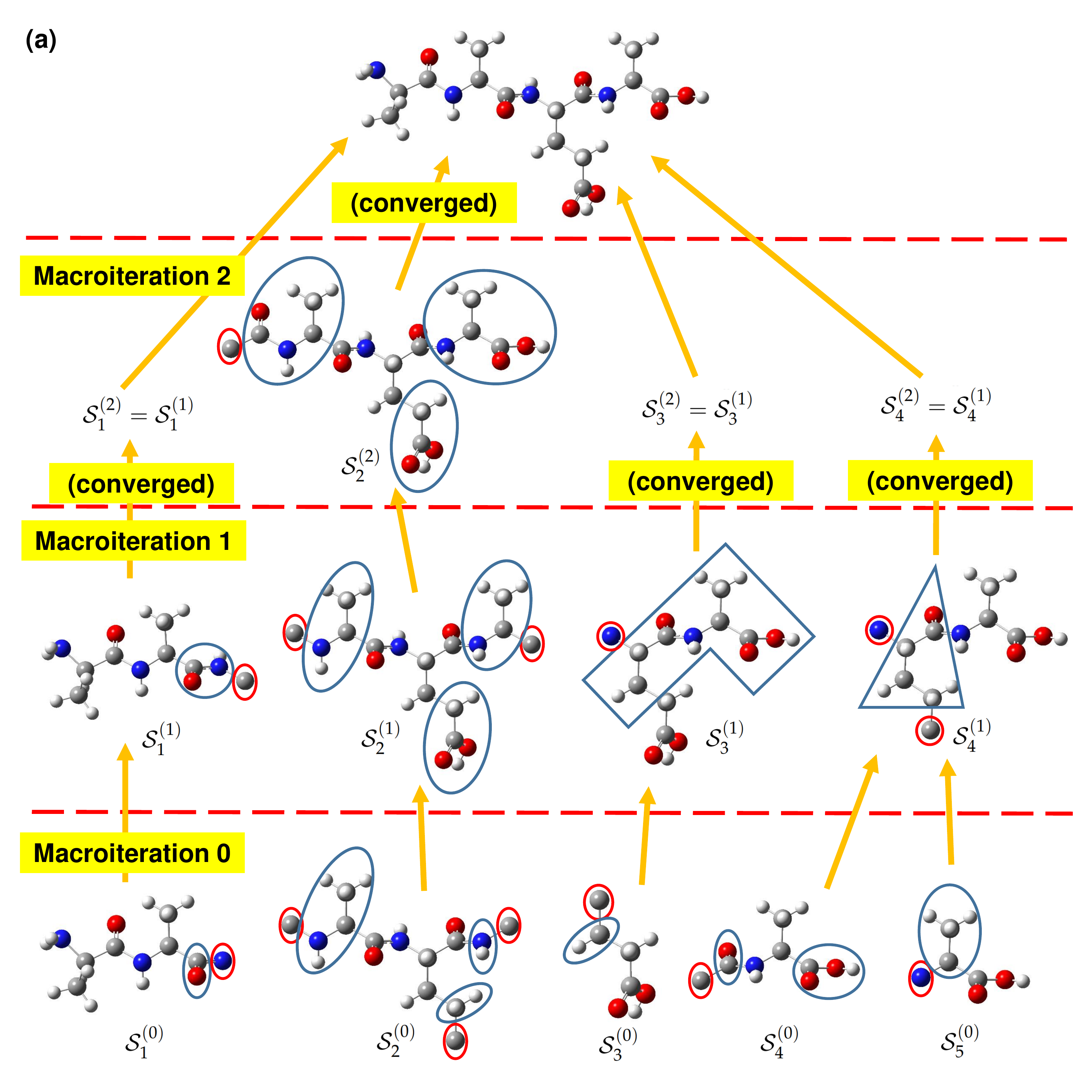}
\includegraphics[width=0.5\textwidth]{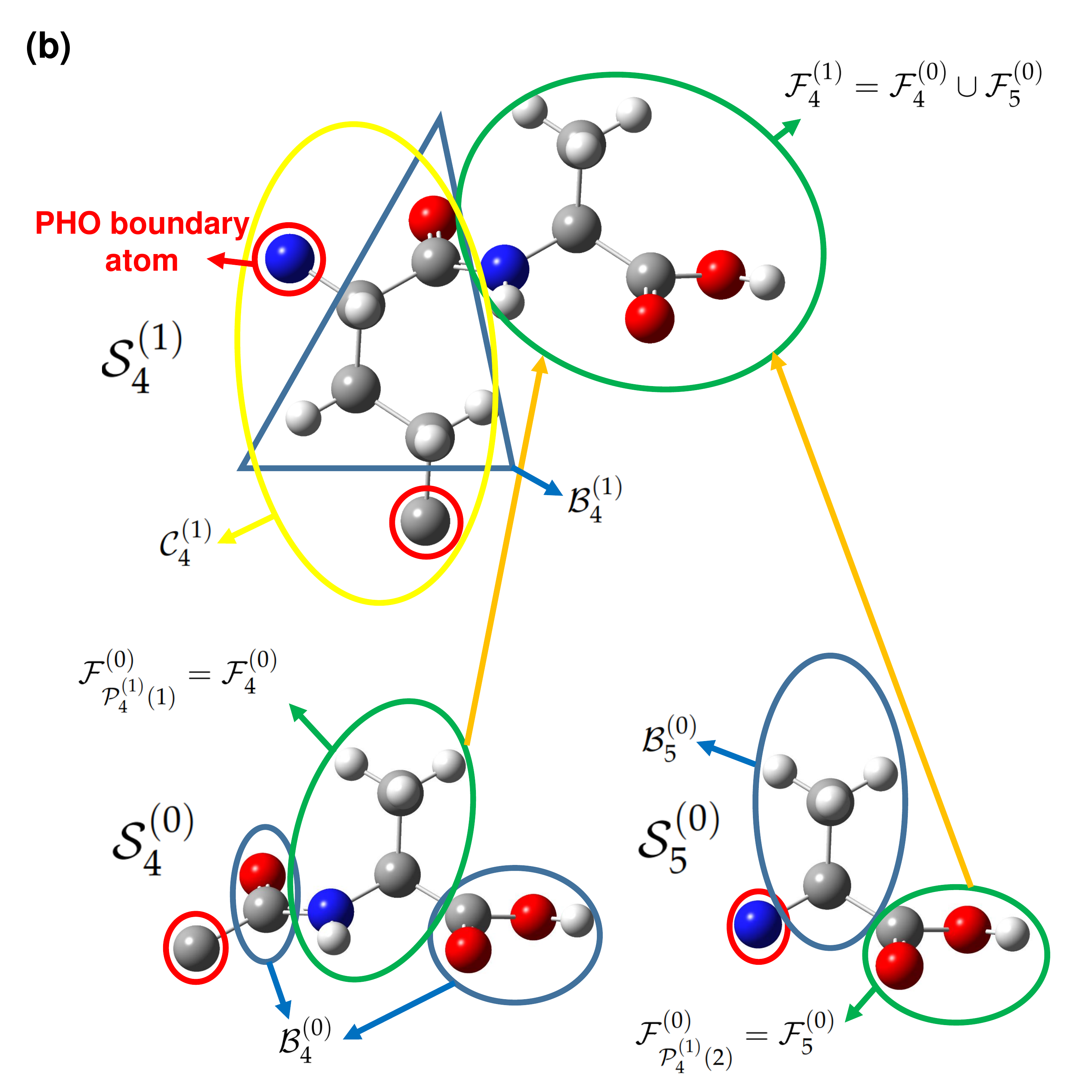}
\caption{Fragmentation of a short peptide. (a) Illustration of the iOI merging process, where the buffer regions and the PHO (projected hybrid orbital)
boundary atoms are designated by blue and red circles, respectively. (b) Definitions of $\mathcal{F}_L^{(\tilde{m})}$, $\mathcal{B}_L^{(\tilde{m})}$,
$\mathcal{C}_L^{(\tilde{m})}$, and $\mathcal{P}_L^{(\tilde{m})}(\cdot)$.}
\label{ala4}
\end{figure}
\item Cap each pFrag with parts of the system that are sufficiently close to the pFrag (see Appendix \ref{autofrag} for the detailed algorithm).
The atoms added in this step to a pFrag $\mathcal{F}_L^{(0)}$
are collectively called buffer $\mathcal{B}_L^{(0)}$; \label{buffering}
\item\label{add_pho} Saturate the dangling bonds to form closed subsystems $\mathcal{S}_L^{(0)}$. Although a natural choice for this purpose is to use hydrogen atoms as link atoms as in our previous work\cite{FLMO1,FLMO3},
it appears more convenient to use the PHO (projected hybrid orbital)\cite{PHO} type of boundary atoms. Specifically, the atoms directly bonded to buffer $\mathcal{B}_L^{(0)}$
     are taken as boundary atoms. Since $\mathcal{B}_L^{(0)}$ has been chosen in such a way that the boundary atoms are always $p$-block element atoms, there always exist
     one $s$ orbital and three $p$ orbitals for each boundary atom, which
     can be hybridized to give four hybrid orbitals (see Appendix \ref{pho} for the modification of the PHO\cite{PHO}).
     Only one of the hybrid orbitals will be available for bonding with one atom in $\mathcal{B}_L^{(0)}$,
     while the remaining three hybrid orbitals (designated as auxiliary PHO) are assigned with suitable occupation numbers and frozen in the SCF calculation.
      The union of $\mathcal{B}_L^{(0)}$ and its boundary atoms is called cap $\mathcal{C}_L^{(0)}$.
     A subsystem is hence the union of a pFrag and its cap, i.e., $\mathcal{S}_L^{(0)} = \mathcal{F}_L^{(0)} \cup \mathcal{C}_L^{(0)}$.
\item\label{SCF_macroiter0} Carry out standard SCF calculation for each subsystem (designated as microiteration), where the auxiliary PHOs are frozen (i.e., projected out of the variational space; see Appendix \ref{pho});
\item Localize the CMOs of each subsystem using, e.g., the Boys localization\cite{BoysLMO}. Note that the auxiliary PHOs are not included in the localization, since they
are not part of the subsystem but only serve to provide a realistic effective potential during the SCF procedure;
\item\label{subsystem} Enlarge successively the subsystems until they are just large enough for expanding the pFLMOs up to a given precision:
\begin{enumerate}
\item\label{pickLMO} For each subsystem, pick up those occupied and virtual LMOs with the L\"owdin populations on its pFrag larger than $\eta_{\mathcal{F}} = 0.1$, leading to pFLMOs $\{\psi^{\mathrm{o, FLMO}}_{\mu}\}_L^{(\tilde{m})}$ and $\{\psi^{\mathrm{v, FLMO}}_{\mu}\}_L^{(\tilde{m})}$
for the occupied and virtual subspaces, respectively;
\item\label{ioimerge} Merge neighboring pFrags of macroiteration $\tilde{m}$ and cap the results to yield larger subsystems. First, construct
$N_{\mathcal{F}}^{(\tilde{m}+1)}$ pFrags for macroiteration $\tilde{m}+1$ such that the $L$th pFrag $\mathcal{F}_L^{(\tilde{m}+1)}$
 is just the union of exactly two proximal pFrags (denoted as $\mathcal{F}_{\mathcal{P}_L^{(\tilde{m}+1)}(1)}^{(\tilde{m})}$ and $\mathcal{F}_{\mathcal{P}_L^{(\tilde{m}+1)}(2)}^{(\tilde{m})}$) of macroiteration $\tilde{m}$, i.e., $\mathcal{F}_L^{(\tilde{m}+1)} = \mathcal{F}_{\mathcal{P}_L^{(\tilde{m}+1)}(1)}^{(\tilde{m})}\bigcup\mathcal{F}_{\mathcal{P}_L^{(\tilde{m}+1)}(2)}^{(\tilde{m})}$. Note that every pFrag of macroiteration $\tilde{m}$ is used exactly one time. The pFrags $\mathcal{F}_{\mathcal{P}_L^{(\tilde{m}+1)}(I)}^{(\tilde{m})}$ ($I=1,2$) are called parents of $\mathcal{F}_L^{(\tilde{m}+1)}$. For example, pFrags $\mathcal{F}_4^{(0)}$ and $\mathcal{F}_5^{(0)}$ in Fig. \ref{ala4} are merged to give pFrag $\mathcal{F}_4^{(1)}$
 (i.e., $\mathcal{P}_4^{(1)}(1) = 4$ and $\mathcal{P}_4^{(1)}(2) = 5$). Note in passing that it is possible to merge more than two pFrags at a time. However,
 the subsystem sizes may increase too rapidly as the macroiteration proceeds. Second, cap the pFrags $\mathcal{F}_L^{(\tilde{m}+1)}$
 in the same way as described in Steps (\ref{buffering}) and (\ref{add_pho}) but with increased buffer radii (see Appendix \ref{merging}), thereby leading to
$N_{\mathcal{F}}^{(\tilde{m}+1)}$ subsystems $\mathcal{S}_L^{(\tilde{m}+1)}$ for macroiteration $\tilde{m}+1$.
 Among the cap atoms added in this step, those that do not belong to  $\mathcal{S}_{\mathcal{P}_L^{(\tilde{m}+1)}(\cdot)}^{(\tilde{m})}$ constitute the incremental cap of subsystem $\mathcal{S}_L^{(\tilde{m}+1)}$, $\Delta\mathcal{C}_L^{(\tilde{m}+1)} = \mathcal{C}_L^{(\tilde{m}+1)} \setminus \bigcup_{I}\mathcal{S}_{\mathcal{P}_L^{(\tilde{m}+1)}(I)}^{(\tilde{m})}$, where $\mathcal{C}_L^{(\tilde{m}+1)}$ is the cap of $\mathcal{S}_L^{(\tilde{m}+1)}$.
 Increase the macroiteration count $\tilde{m}$ by 1.

 The above merging scheme improves the quality of the pFLMOs by using more AOs and improved effective potential $V_{eff}(\boldsymbol{r})$. Yet, there exist
 several exceptions to these rules:
  \begin{enumerate}
  \item It is obvious that, if the AOs of subsystem $\mathcal{S}_L^{(\tilde{m})}$ are already sufficient for accurately expanding the pFLMOs, $\mathcal{S}_L^{(\tilde{m})}$ need not be further enlarged. This can readily be checked by the total L\"owdin population, $(\Delta_e)_L^{(\tilde{m})}$, of the occupied pFLMOs $\{\psi^{\mathrm{o, FLMO}}_{\mu}\}_L^{(\tilde{m})}$ on the incremental cap $\Delta\mathcal{C}_L^{(\tilde{m})}$,
  \begin{eqnarray}
  (\Delta_e)_L^{(\tilde{m})}&= & \sum_{\nu\in\Delta\mathcal{C}_L^{(\tilde{m})}} \left((\mathcal{S}_L^{(\tilde{m})})^{1/2}\mathbf{\tilde{D}}_L^{(\tilde{m})}(\mathcal{S}_L^{(\tilde{m})})^{1/2}\right)_{\nu\nu} \nonumber\\
  & = & \sum_{\nu\in\Delta\mathcal{C}_L^{(\tilde{m})}}\sum_{i\in\{\psi^{\mathrm{o, FLMO}}_{\mu}\}_L^{(\tilde{m})}} \left|\left((\mathcal{S}_L^{(\tilde{m})})^{1/2}\mathbf{C}_L^{(\tilde{m})}\right)_{\nu i}\right|^2,\\
  (\mathbf{\tilde{D}}_L^{(\tilde{m})})_{\nu\kappa} &=& \sum_{i\in\{\psi^{\mathrm{o, FLMO}}_{\mu}\}_L^{(\tilde{m})}}(\mathbf{C}_L^{(\tilde{m})})_{\nu i}(\mathbf{C}_L^{(\tilde{m})})_{\kappa i},
  \end{eqnarray}
  where $\mathcal{S}_L^{(\tilde{m})}$ and $\mathbf{C}_L^{(\tilde{m})}$ are the AO overlap matrix and the LMO coefficients of $\mathcal{F}_L^{(\tilde{m})}$, respectively.
  If $(\Delta_e)_L^{(\tilde{m})}<\eta_{\mathrm{tail}}=0.1$, subsystem $\mathcal{S}_L^{(\tilde{m})}$ is considered to be converged and remains unchanged in subsequent macroiterations.
  Such converged subsystems have just enough sizes. For example, subsystems $\mathcal{S}_1^{(1)}$, $\mathcal{S}_3^{(1)}$ and $\mathcal{S}_4^{(1)}$ in Fig. \ref{ala4} are converged at macroiteration 1, but $\mathcal{S}_2^{(1)}$ is not yet so and has to be enlarged once more to give $\mathcal{S}_2^{(2)}$.

  \item When the total number of nonconvergent subsystems is odd or when a pFrag at macroiteration $\tilde{m}$ is spatially distant from the remaining ones, it is impossible to merge all pFrags in a pairwise fashion. In these cases, some pFrags of macroiteration $\tilde{m}+1$ may contain only one pFrag of macroiteration $\tilde{m}$, i.e., $\mathcal{F}_L^{(\tilde{m}+1)} = \mathcal{F}_{\mathcal{P}_L^{(\tilde{m}+1)}(1)}^{(\tilde{m})}$. However, $\mathcal{S}_L^{(\tilde{m}+1)}$ is still larger than $\mathcal{S}_{\mathcal{P}_L^{(\tilde{m}+1)}(1)}^{(\tilde{m})}$ since $\mathcal{C}_L^{(\tilde{m}+1)}$ is larger than $\mathcal{C}_{\mathcal{P}_L^{(\tilde{m}+1)}(1)}^{(\tilde{m})}$. For example, at macroiteration 0, there are five pFrags for the example shown in Fig. \ref{ala4}. For load balance,  $\mathcal{F}_2^{(0)}$ is chosen as the pFrag that will not undergo merging for $\mathcal{S}_2^{(0)}$ is the largest of the five subsystems. This separates $\mathcal{F}_1^{(0)}$ from the other three pFrags, such that $\mathcal{F}_1^{(0)}$ and $\mathcal{F}_3^{(0)}$ are also directly carried over to the next macroiteration. Then, only $\mathcal{F}_4^{(0)}$ and $\mathcal{F}_5^{(0)}$ are merged (for details, see Appendix \ref{merging}).
  \item When all subsystems have been converged (large enough), merge them together and stop the macroiterations. If all (nonconvegent) subsystems will be merged together at a macroiteration, also stop the macroiterations. The iOI procedure is then finished by performing a small number of global SCF iterations (see Step (\ref{global})).
  \end{enumerate}
\item Use the pFLMOs of macroiteration $\tilde{m}-1$ as initial guess for macroiteration $\tilde{m}$. Specifically, find all subsystems $\mathcal{S}_{L_i^{(\tilde{m}-1)}}^{(\tilde{m}-1)}$ ($i\in[1,\mathcal{M}_L^{(\tilde{m})}])$ whose pFrags overlap with subsystem $\mathcal{S}_L^{(\tilde{m})}$, i.e., find $\{L_i^{(\tilde{m}-1)}\}$ with $\mathcal{S}_L^{(\tilde{m})} \cap \mathcal{F}_{L_i^{(\tilde{m}-1)}}^{(\tilde{m}-1)} \neq \varnothing$. For example, in Fig. \ref{ala4}, we have $\mathcal{M}_1^{(1)} = 2$, $1_1^{(0)} = 1$ and $1_2^{(0)} = 2$, since $\mathcal{S}_1^{(1)}$ overlaps with $\mathcal{F}_1^{(0)}$ and $\mathcal{F}_2^{(0)}$ but not with $\mathcal{F}_3^{(0)}$, $\mathcal{F}_4^{(0)}$ or $\mathcal{F}_5^{(0)}$. Note that $\mathcal{P}_L^{(\tilde{m})}(\cdot)\in \{L_i^{(\tilde{m}-1)}\}$, since $\mathcal{F}_{\mathcal{P}_L^{(\tilde{m})}(\cdot)}^{(\tilde{m}-1)} \subseteq \mathcal{F}_L^{(\tilde{m})}$, which implies necessarily that $\mathcal{F}_{\mathcal{P}_L^{(\tilde{m})}(\cdot)}^{(\tilde{m}-1)}$ overlaps with $\mathcal{S}_L^{(\tilde{m})}$. The pFLMOs of
    subsystems $\{\mathcal{S}_{L_i^{(\tilde{m}-1)}}^{(\tilde{m}-1)}\}$ are then projected onto subsystem $\mathcal{S}_L^{(\tilde{m})}$, i.e.,
  \begin{equation}
  (\mathbf{C}_L^{(\tilde{m})})_{\kappa p} = \sum_{\nu\in \mathcal{S}_L^{(\tilde{m})}}((\mathcal{S}_L^{(\tilde{m})})^{-1})_{\kappa\nu} (\mathcal{S}_{L^{(\tilde{m})}, L_i^{(\tilde{m}-1)}}\mathbf{C}_{L_i^{(\tilde{m}-1)}}^{(\tilde{m}-1)})_{\nu p},\nonumber
  \end{equation}
  \begin{equation}
  p\in \{\psi^{\mathrm{o, FLMO}}_{\mu}\}_{L_i^{(\tilde{m}-1)}}^{(\tilde{m}-1)}\cup\{\psi^{\mathrm{v, FLMO}}_{\mu}\}_{L_i^{(\tilde{m}-1)}}^{(\tilde{m}-1)}, \label{projection}
  \end{equation}
  where $\mathcal{S}_L^{(\tilde{m})}$ and $\mathcal{S}_{L^{(\tilde{m})}, L_i^{(\tilde{m}-1)}}$ are the AO overlap matrix of subsystem $\mathcal{S}_L^{(\tilde{m})}$ and the rectangular overlap matrix between the AOs of $\mathcal{S}_L^{(\tilde{m})}$ and $\mathcal{S}_{L_i^{(\tilde{m}-1)}}^{(\tilde{m}-1)}$, respectively. Physically, the projection Eq. \eqref{projection} can be interpreted as finding the pFLMOs $\{(\mathbf{C}_L^{(\tilde{m})})_p\}$ that resemble the input pFLMOs $\{(\mathbf{C}_{L_i^{(\tilde{m}-1)}}^{(\tilde{m}-1)})_p\}$ in the least-squares sense, i.e. the $(\mathbf{C}_L^{(\tilde{m})})_p$ defined in Eq. \eqref{projection} minimizes the differential inner product $\langle(\mathbf{C}_L^{(\tilde{m})})_p-(\mathbf{C}_{L_i^{(\tilde{m}-1)}}^{(\tilde{m}-1)})_p|(\mathbf{C}_L^{(\tilde{m})})_p-(\mathbf{C}_{L_i^{(\tilde{m}-1)}}^{(\tilde{m}-1)})_p\rangle$ subject to the constraint that $\{(\mathbf{C}_L^{(\tilde{m})})_p\}$ forms an orthonormal set\cite{PAO}. As mentioned above, the parental subsystems of $\mathcal{S}_L^{(\tilde{m})}$, $\mathcal{S}_{\mathcal{P}_L^{(\tilde{m})}(\cdot)}^{(\tilde{m}-1)}$, must be members of subsystems $\{\mathcal{S}_{L_i^{(\tilde{m}-1)}}^{(\tilde{m}-1)}\}$ since the pFrags of the former overlap with $\mathcal{S}_L^{(\tilde{m})}$ by construction. Of all pFLMOs from $\{\mathcal{S}_{L_i^{(\tilde{m}-1)}}^{(\tilde{m}-1)}\}$, those from $\mathcal{S}_{\mathcal{P}_L^{(\tilde{m})}(\cdot)}^{(\tilde{m}-1)}$ are denoted as central pFLMOs of $\mathcal{S}_L^{(\tilde{m})}$, i.e., $\{\xi_{\mu}^{\mathrm{o}}\}_L^{(\tilde{m})}$ and $\{\xi_{\mu}^{\mathrm{v}}\}_L^{(\tilde{m})}$,
  whereas the remaining ones are called  buffer pFLMOs of $\mathcal{S}_L^{(\tilde{m})}$. Note that, in the presence of PHO boundary atoms, the contributions of the auxiliary PHOs should be projected out from the central pFLMOs;
\item\label{LELD} For each subsystem, perform ``local elimination of linear dependence'' (LELD) in $\{\xi_{\mu}^{\mathrm{o}}\}_L^{(\tilde{m})}$ and $\{\xi_{\mu}^{\mathrm{v}}\}_L^{(\tilde{m})}$.
In the original LELD algorithm,\cite{ACR-FLMO} we eliminate the $\xi^{\mathrm{o}}_{\mu}$ that has the largest overlap with the eigenvector $\{v_p\}$ of the overlap matrix
$\langle\xi^{\mathrm{o}}_{\mu}|\xi^{\mathrm{o}}_{\nu}\rangle$ with the smallest eigenvalue. In case of degeneracy,
the $\xi_{\mu}^{\mathrm{o}}$ with the largest orbital spread $O[\xi_{\mu}^{\mathrm{o}}] = 2(\langle\xi_{\mu}^{\mathrm{o}}|\widehat{r^2}|\xi_{\mu}^{\mathrm{o}}\rangle - |\langle\xi_{\mu}^{\mathrm{o}}|\hat{\mathbf{r}}|\xi_{\mu}^{\mathrm{o}}\rangle|^2)$ is first eliminated. The overlap matrix of the survival pFLMOs is then re-diagonalized, and the process is repeated until the number of remaining occupied and virtual orbitals match those expected from the
numbers of electrons and basis functions of the subsystem. A na\"ive implementation of this algorithm requires diagonalizing the global pFLMO overlap matrix $N_{\mathrm{red}}$ times (NB: $N_{\mathrm{red}}$ is the number of redundant pFLMOs), which scales as $\mathcal{O}(N_{\mathrm{pFLMO}}^3N_{\mathrm{red}})$. However, the pFLMO overlap matrix has a blocked structure, so that its diagonalization can be approximately replaced by the diagonalization of $O(N_{\mathrm{pFLMO}})$ small matrices, each of dimension $O(1)$. The revised LELD algorithm thus scales linearly with respect to system size (the detailed algorithm will be reported in a separate paper). The linearly independent occupied pFLMOs are then L\"owdin orthonormalized.
The virtual pFLMOs are treated in the same way after projecting out the orthonormal occupied pFLMOs;
\item\label{SCF_macroiterN} Do subsystem SCF calculations.
To retain the locality of the LMOs throughout the SCF calculation, we here use the bottom-up, least-change block-diagonalization\cite{ACR-FLMO} instead of the conventional full diagonalization
of the Fock matrix $\mathbf{F}$ represented in the pFLMO basis and written in block form,
\begin{equation}
\mathbf{F}=\begin{pmatrix} \mathbf{F}_{\mathrm{\tilde{o}\tilde{o}}} & \mathbf{F}_{\mathrm{\tilde{o}v}} \\
\mathbf{F}_{\mathrm{v\tilde{o}}} & \mathbf{F}_{\mathrm{vv}} \end{pmatrix},\label{F:CMO}
\end{equation}
where the subscript $\mathrm{\tilde{o}}$ emphasizes that the occupied buffer pFLMOs and the auxiliary PHOs are all frozen.
While freezing the auxiliary PHOs is required by the PHO method itself\cite{PHO}, the rationale for freezing also the occupied buffer pFLMOs of subsystem $\mathcal{S}_L^{(\tilde{m})}$
is that they come from the pFrags $\mathcal{F}_{L_i^{(\tilde{m}-1)}}^{(\tilde{m}-1)}$ of neighboring subsystems, where they experience a very realistic environment provided by the caps $\mathcal{C}_{L_i^{(\tilde{m}-1)}}^{(\tilde{m}-1)}$ of these subsystems. In contrast, as the cap of $\mathcal{S}_L^{(\tilde{m})}$, the occupied buffer pFLMOs experience an inaccurate effective potential. Therefore, their relaxation would actually worsen themselves and in turn also worsen
the central pFLMOs of $\mathcal{S}_L^{(\tilde{m})}$.
Following the same mathematics as we go from the matrix Dirac equation to the exact two-component relativistic theory\cite{X2C2005,X2C2009},
we can define the following unitary transformation\cite{LiuMP}
\begin{eqnarray}
\mathbf{U}&=&\mathbf{D}\mathbf{P},\label{Umat}\\
\mathbf{D}&=&\begin{pmatrix} \mathbf{I}_{\mathrm{\tilde{o}\tilde{o}}} & -\mathbf{X}^\dag \\
             \mathbf{X} & \mathbf{I}_{\mathrm{vv}} \end{pmatrix},\label{DecMat}\\
\mathbf{P}&=&\begin{pmatrix} \tilde{\mathcal{S}}_{\mathrm{\tilde{o}\tilde{o}}}^{-1/2} & \mathbf{0} \\
             \mathbf{0} & \tilde{\mathcal{S}}_{\mathrm{vv}}^{-1/2} \end{pmatrix},\label{NormMat}\\
\tilde{\mathcal{S}}_{\mathrm{\tilde{o}\tilde{o}}}&=&\mathbf{I}_{\mathrm{\tilde{o}\tilde{o}}}+\mathbf{X}^\dag\mathbf{X},\quad \tilde{\mathcal{S}}_{\mathrm{vv}}=\mathbf{I}_{\mathrm{vv}}+\mathbf{X}\mathbf{X}^\dag,
\end{eqnarray}
which block-diagonalizes the Fock matrix \eqref{F:CMO} as
\begin{eqnarray}
\bar{\mathbf{F}} &=& \mathbf{U}^\dag \mathbf{F} \mathbf{U} =
\begin{pmatrix} \bar{\mathbf{F}}_{\mathrm{\tilde{o}\tilde{o}}} & \mathbf{0} \\
\mathbf{0} & \bar{\mathbf{F}}_{\mathrm{vv}} \end{pmatrix},\label{F:LMO}\\
\bar{\mathbf{F}}_{\mathrm{\tilde{o}\tilde{o}}}&=&\tilde{\mathcal{S}}_{\mathrm{\tilde{o}\tilde{o}}}^{-1/2}\tilde{\mathbf{F}}_{\mathrm{\tilde{o}\tilde{o}}}\tilde{\mathcal{S}}_{\mathrm{\tilde{o}\tilde{o}}}^{-1/2},\\
\bar{\mathbf{F}}_{\mathrm{vv}}&=&\tilde{\mathcal{S}}_{\mathrm{vv}}^{-1/2}\tilde{\mathbf{F}}_{\mathrm{vv}}\tilde{\mathcal{S}}_{\mathrm{vv}}^{-1/2},\\
\tilde{\mathbf{F}}_{\mathrm{\tilde{o}\tilde{o}}}&=&\mathbf{F}_{\mathrm{\tilde{o}\tilde{o}}}+\mathbf{F}_{\mathrm{\tilde{o}v}}\mathbf{X}+\mathbf{X}^\dag\mathbf{F}_{\mathrm{v\tilde{o}}}+\mathbf{X}^\dag\mathbf{F}_{\mathrm{vv}}\mathbf{X},\label{FLMOoo}\\
\tilde{\mathbf{F}}_{\mathrm{vv}}&=&\mathbf{F}_{\mathrm{vv}}-\mathbf{X}\mathbf{F}_{\mathrm{\tilde{o}v}}-\mathbf{F}_{\mathrm{v\tilde{o}}}\mathbf{X}^\dag+\mathbf{X}\mathbf{F}_{\mathrm{\tilde{o}\tilde{o}}}\mathbf{X}^\dag,\label{FLMOvv}\\
\tilde{\mathbf{F}}_{\mathrm{v\tilde{o}}}&=&\mathbf{F}_{\mathrm{v\tilde{o}}}-\mathbf{X}\mathbf{F}_{\mathrm{\tilde{o}\tilde{o}}}+\mathbf{F}_{\mathrm{vv}}\mathbf{X}
-\mathbf{X}\mathbf{F}_{\mathrm{\tilde{o}v}}\mathbf{X}=\tilde{\mathbf{F}}_{\mathrm{\tilde{o}v}}^\dag=\mathbf{0}.\label{eq:Xeq}
\end{eqnarray}
Once the pFLMO Fock matrix \eqref{F:CMO} is available, the decoupling condition \eqref{eq:Xeq} for $\mathbf{X}$ can be solved iteratively with the Jacobi sweep method.
The $\mathbf{U}$ matrix \eqref{Umat} is just the coefficient matrix of the LMOs in the pFLMO basis (which is further expanded in the AO basis).

\item go to Step (\ref{pickLMO}).
\end{enumerate}
\item\label{global} Do global SCF. Since the target here is to have fully converged occupied and virtual LMOs of the whole system, this global SCF or last macroiteration is necessary.
\begin{enumerate}
\item Gather the central pFLMOs of all subsystems and generate linearly independent, orthonormal pFLMOs in the same way as described in Step (\ref{LELD});
\item Like subsystem calculations, the bottom-up, least-change block-diagonalization of the Fock matrix is used. At each iteration the dimension of the Fock matrix $\mathbf{F}$ \eqref{F:CMO}
can be reduced by freezing the already converged LMOs:
 (1) occupied LMOs $i$ that satisfy $\max_{a} |F_{ia}| < \eta_{\mathrm{Fov}} = 10^{-4}$ are frozen, where $a$ runs over all virtual LMOs; (2) virtual LMOs $a$ that satisfy $\max_{i} |F_{ia}| < \eta_{\mathrm{Fov}}$ are frozen, where $i$ runs over the active occupied LMOs only. Since the off-diagonal elements
 $\{F_{ia}\}$ tend to diminish along the SCF iterations, usually more and more LMOs can be frozen.
 However, some frozen LMOs of an iteration may no longer be frozen in the subsequent iteration, such that the number of frozen LMOs may occasionally decrease.
\end{enumerate}
\end{enumerate}

Some further remarks are in order.
\begin{enumerate}[(I)]
\item
If the global SCF calculation (the last macroiteration of iOI) with a sufficiently small threshold $\eta_{\mathrm{Fov}}$ for orbital freezing
is iterated till convergence, both the total energy and the wave function (LMOs) will agree completely with those by the conventional SCF calculation. Therefore, iOI is an exact method.
Even in this case, for sufficiently large systems, iOI can still achieve considerable speedup over the conventional SCF algorithm because
the gain in efficiency due to the much reduced number of global SCF iterations overcompensates for the cost of the subsystem calculations, which need not be fully converged
(e.g., to $10^{-3}$ a.u. for the energy change and $10^{-2}$ a.u. for the maximum change of the density matrix elements) and can be made embarassingly parallel.
Both occupied and virtual LMOs of the global system are obtained for free at the end of the global SCF calculation.
This is important for the top-down, iterative localization of the CMOs of large systems is very costly or even impossible.
Moreover, iOI offers an effective means to converge to a user-specified configuration by assigning proper occupations of the pFrags\cite{FLMO3}
when the global system has more than one stable SCF solutions (which is very common in, e.g., antiferromagnetically coupled systems).
\item
iOI shares the same spirit as iCI\cite{SDS,iCI} for solving the full CI problem: (a) the orbitals of the small and larger subsystems in iOI correspond, respectively, to the reference functions and singly and doubly excited configurations in iCI; (b)
the subsystem and global SCF calculations correspond to the microiterations of iCI for relaxing the contraction coefficients of the CI vectors; (c) both iOI and iCI
can be viewed as a particular sequential, exact (partial) diagonalization of a large matrix, by getting first the roots of one portion of the
matrix and then those of an enlarged portion, until the full matrix has been sampled; (d) from the mathematical point of view, the only distinction between iOI and iCI lies in that the Fockian in the former depends on the unknowns whereas
the Hamiltonian in the latter does not; (d) if only the active part of an exceedingly large systems is of interest, some selection of the orbitals localized on neighboring buffers
can be introduced to iOI, just like the selected iCI\cite{iCIPT2,iCIPT2New}.
\end{enumerate}

\section{Results and discussion}\label{Results}
As showcases, we herein report iOI calculations of a series of double-stranded DNA molecules of various lengths, hereafter termed as DNA$_n (n=1,2,4,8,16)$,
where one strand is composed of $n$ consecutive adenosines (A) and the other is composed of $n$ thymidines (T). All calculations were performed with the BDF program package\cite{BDF1,BDF2,BDF3,BDFECC,BDFrev2020}
on a server equipped with 16 Intel(R) Xeon(R) CPU E5-2640 v3 @ 2.60GHz cores. The B3LYP functional\cite{Becke93,B3LYP} and the def2-SV(P)\cite{def2}
basis set were used throughout. Unless otherwise noted, all SCF calculations of the global system (including the exact SCF and the last macroiteration of iOI) are converged to the following criteria:
$< 10^{-6} \mathrm{~a.u.}$ for the energy change and $< 10^{-4} \mathrm{~a.u.}$ for the maximum change of density matrix elements.
The subsystem calculations of DNA$_n$ were embarrassingly parallelized over $N_{\mathrm{procs}}=\max(2n,16)$ processes,
each with $16/N_{\mathrm{procs}}$ OpenMP threads, while the global system calculations were performed using 16 OpenMP threads.

The first molecule to be studied is the DNA$_{16}$ molecule (1052 atoms, 10024 basis functions). The statistics of the iOI macroiterations is documented in
Table \ref{dna16_subsys}. It is first seen that iOI takes 4 macroiterations to converge (which, of course, depends on
the particular design of subsystem sizes). The number of subsystems decreases along the macroiteration as a result of subsystem merging,
which is accompanied by an increased average size of the subsystems. Since all subsystems are converged at macroiteration 2, all of them are merged together at macroiteration 3
for final SCF iterations of the whole system. Note that the subsystems at macroiteration 2 are still much smaller in size than the whole system,
thereby explaining the shorter timing of macroiteration 2 than that of the final macroiteration, needless to say the first two macroiterations.

\begin{table}
\centering
\caption{Statistics of the iOI macroiterations for DNA$_{16}$.}
\begin{tabular}{l|cccc}
\hline\hline
Macroiteration & 0 & 1 & 2 & 3$^{a}$ \\
\hline
No. of subsystems & 96	&48	&38	&1 \\
No. of converged subsystems & 0	&21	&38	&1\\
Minimal No. of atoms in a subsystem & 7	&36	&36	&1052\\
Maximum No. of atoms in a subsystem & 56	&114	&195	&1052\\
Average No. of atoms in a subsystem & 26	&76	&101	&1052\\
Wall time (s) & 1441 	&2476 	&5845 	&75683 \\
\hline\hline
\end{tabular} \label{dna16_subsys}\\
\footnotesize{$^{a}$ For the whole system.\\}
\end{table}

\begin{figure}
\centering
\begin{minipage}[c]{0.45\textwidth}
\centering
\includegraphics[width=\textwidth]{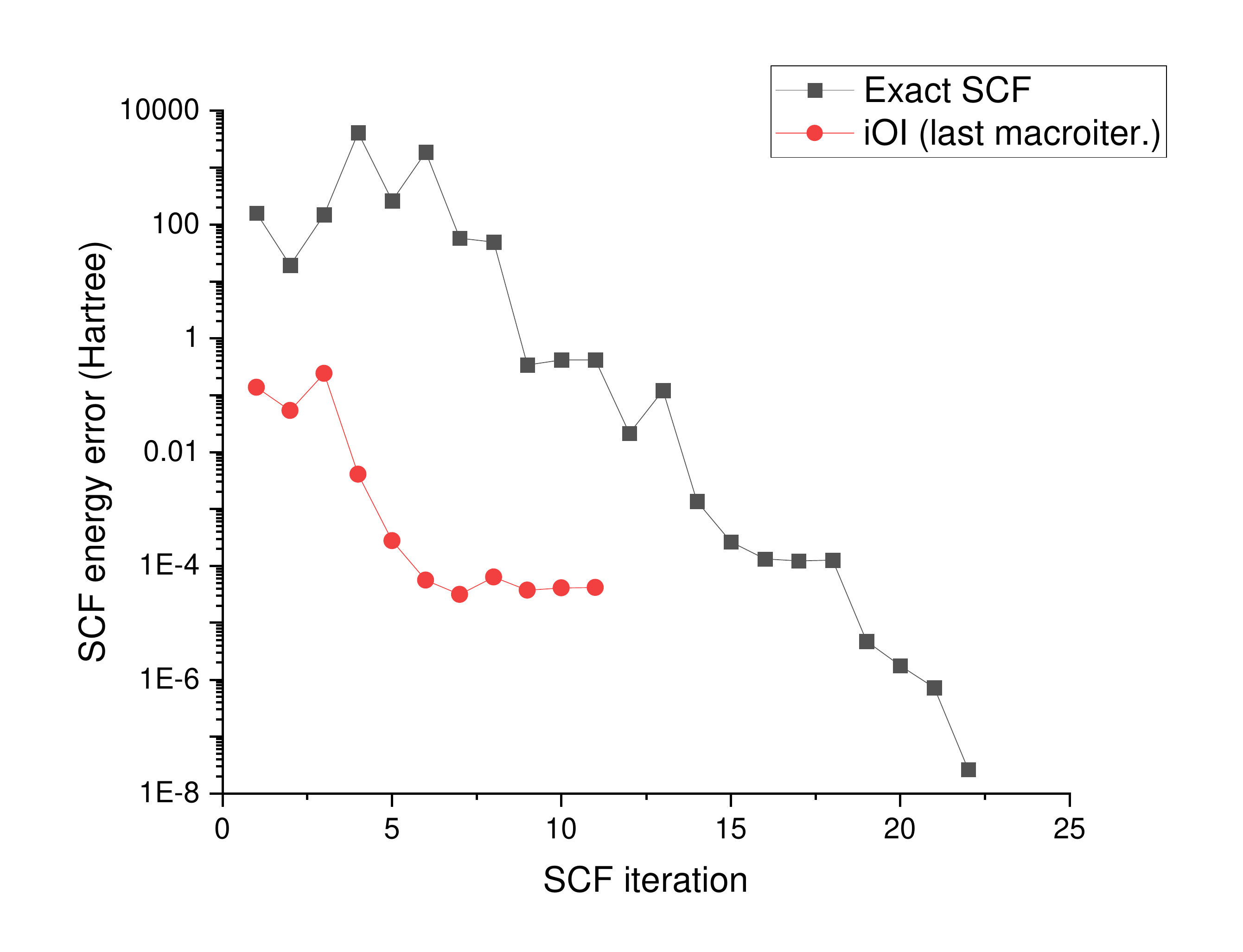}
\includegraphics[width=\textwidth]{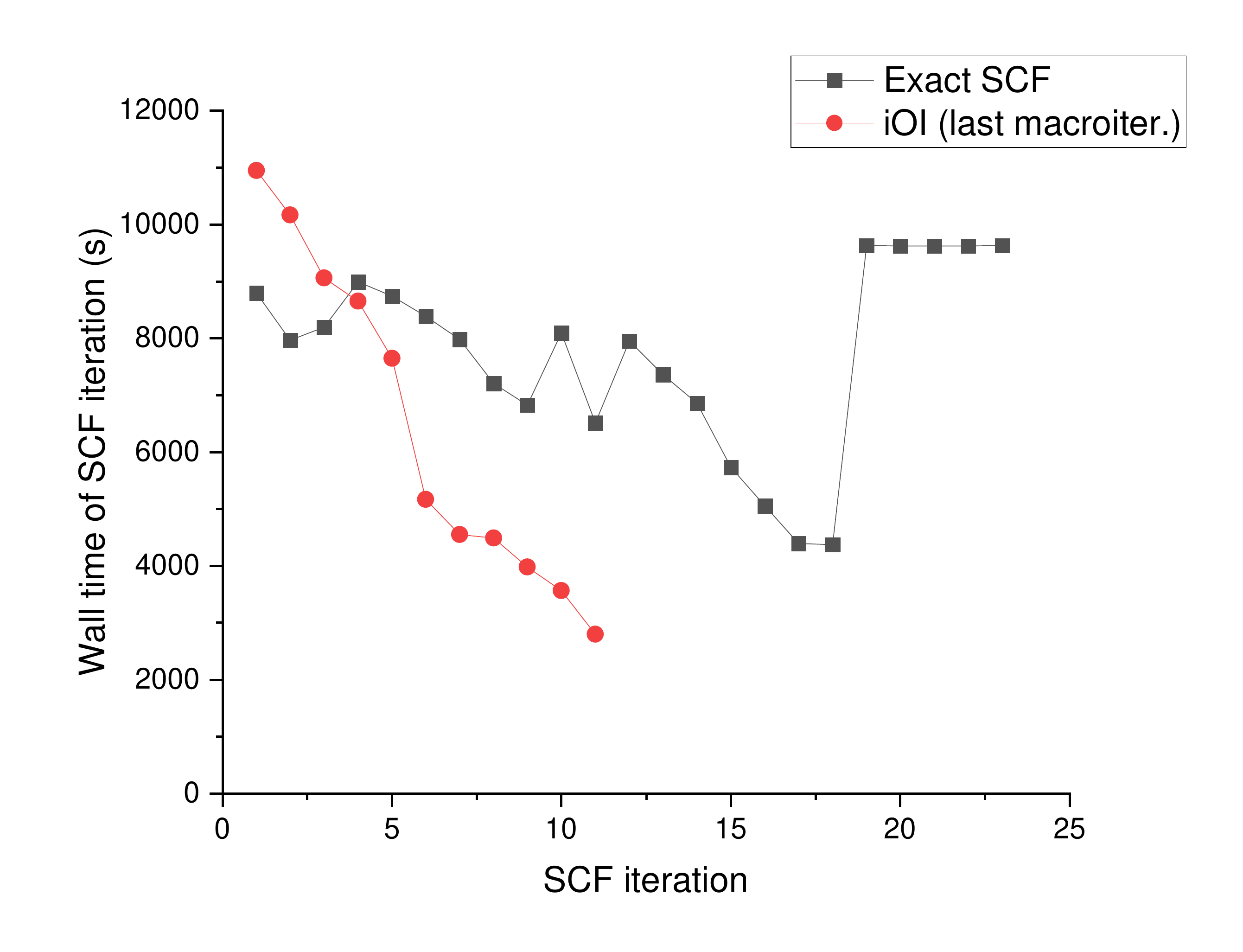} 
\end{minipage}
\begin{minipage}[c]{0.45\textwidth}
\centering
\includegraphics[width=\textwidth]{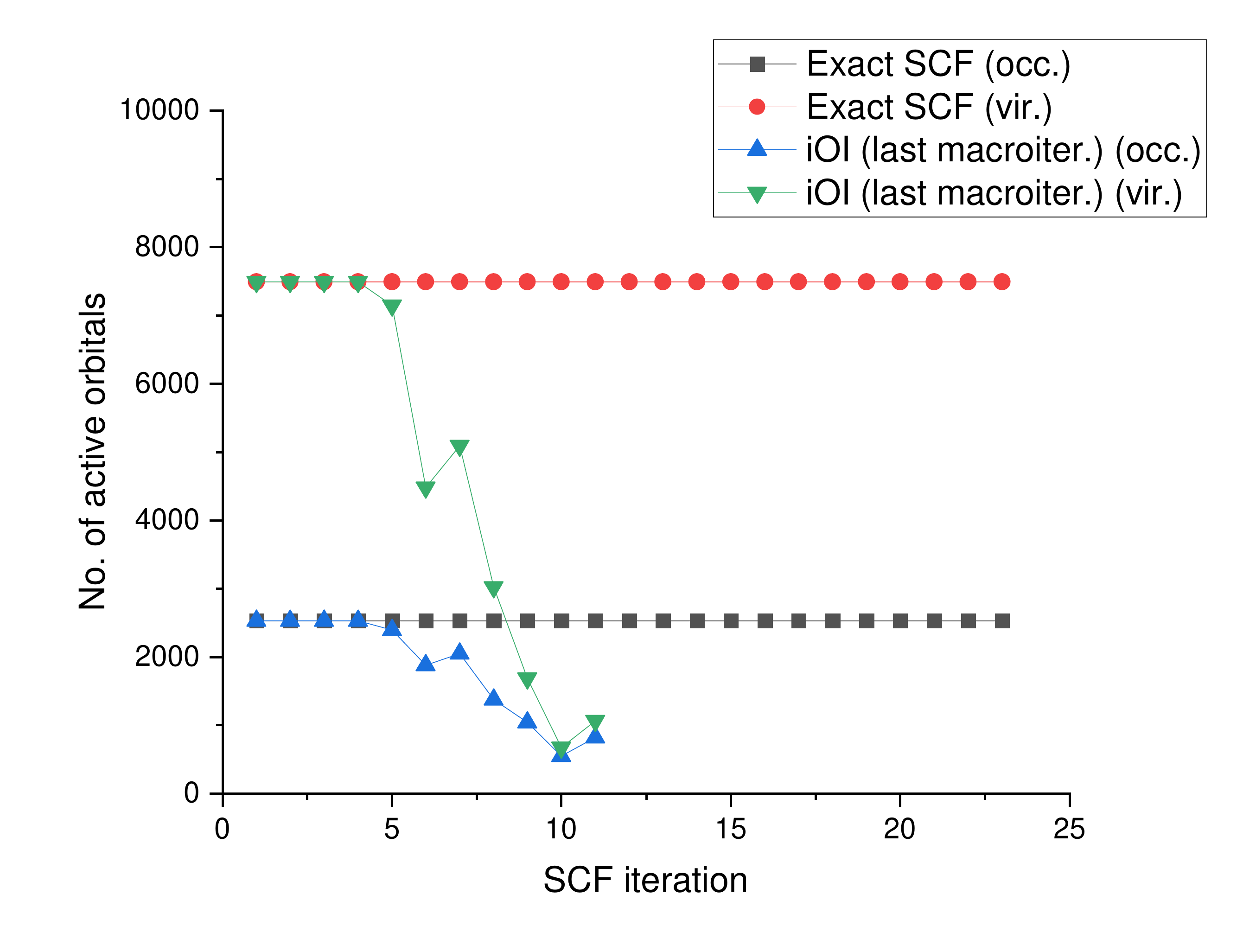}
\includegraphics[width=\textwidth]{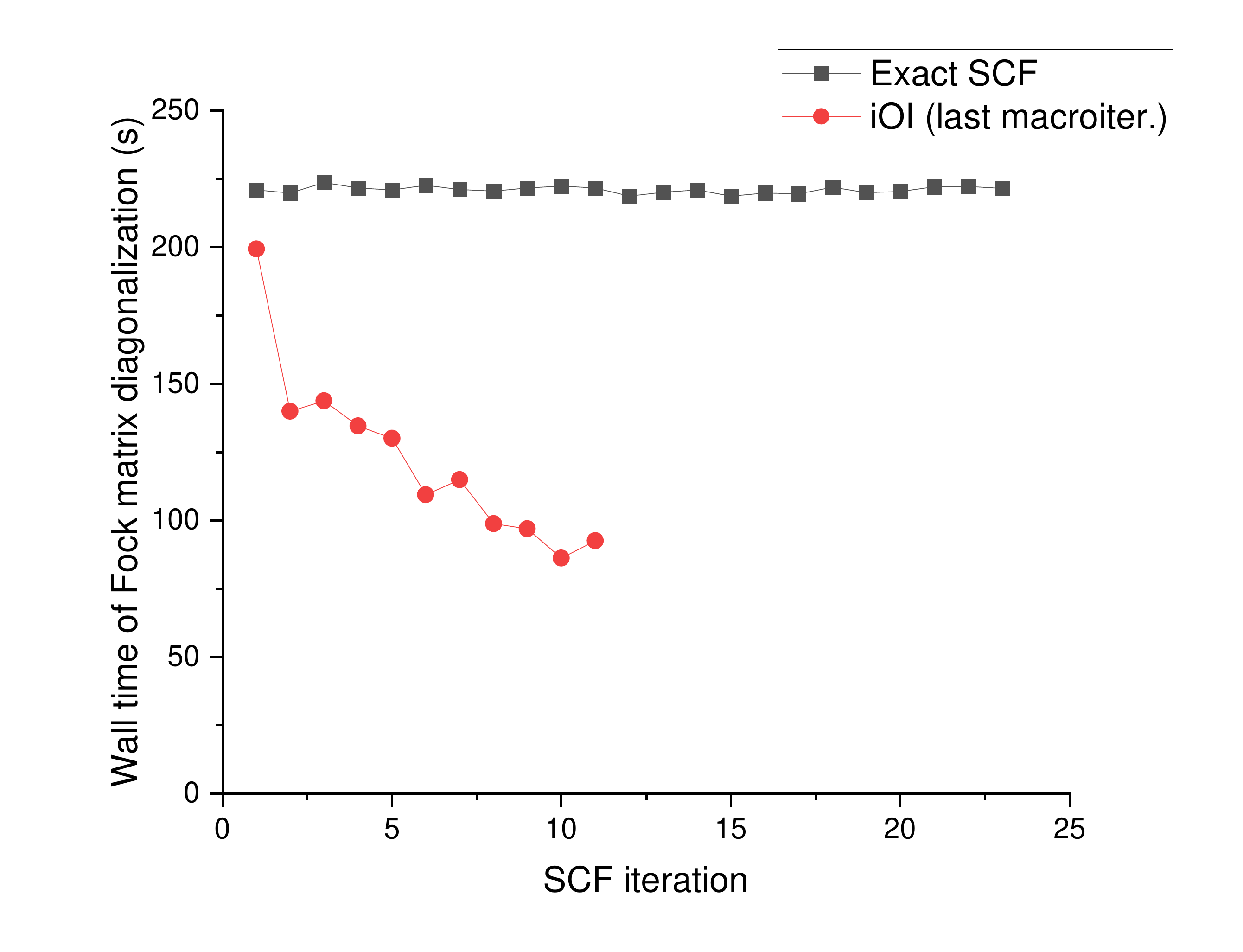}
\end{minipage}
\caption{Comparison of the final iOI macroiteration and exact SCF calculations of DNA$_{16}$.
Top left: error per atom of the iOI energy compared with the exact SCF energy; top right: number of active occupied and virtual orbitals; bottom left: wall time of each SCF iteration; bottom right:
wall time of the Fock matrix (block-)diagonalization step of each SCF iteration.}
\label{DNA16_scf}
\end{figure}

The details of the final macroiteration of iOI is shown in Fig. \ref{DNA16_scf} to compare with the exact SCF calculation starting from the superposition of atomic densities.
It can first be seen that the last macroiteration of iOI converges within 11 iterations, whereas the exact SCF calculation takes 23 iterations to converge.
The first three iOI macroiterations thus generate a much better guess than the atomic guess and consequently save 12 iterations,
 at a cost (9762 s) that corresponds roughly to one global SCF iteration (Fig. \ref{DNA16_scf}, bottom left).
The iOI total energy is in error of $4\times 10^{-5}$ a.u. ($4\times 10^{-8}$ a.u. per atom)
compared to the exact SCF energy (Fig. \ref{DNA16_scf}, top left), which is a result of freezing the nearly converged LMOs.
The number of active LMOs decreases rapidly as the SCF iterations proceed, approaching eventually to only one fifth of the total number of LMOs (Fig. \ref{DNA16_scf}, top right),
which results in a marked speedup of the Fock matrix block-diagonalization over the exact diagonalization (Fig. \ref{DNA16_scf}, bottom right).
It is interesting to note that the proportion of frozen virtual LMOs is much higher than that of frozen occupied LMOs (Fig. \ref{DNA16_scf}, top right),
which can be rationalized by the fact that there are typically many more Rydberg orbitals than core orbitals with extended basis sets.

\begin{figure}
\centering
\begin{minipage}[c]{0.45\textwidth}
\centering
\includegraphics[width=\textwidth]{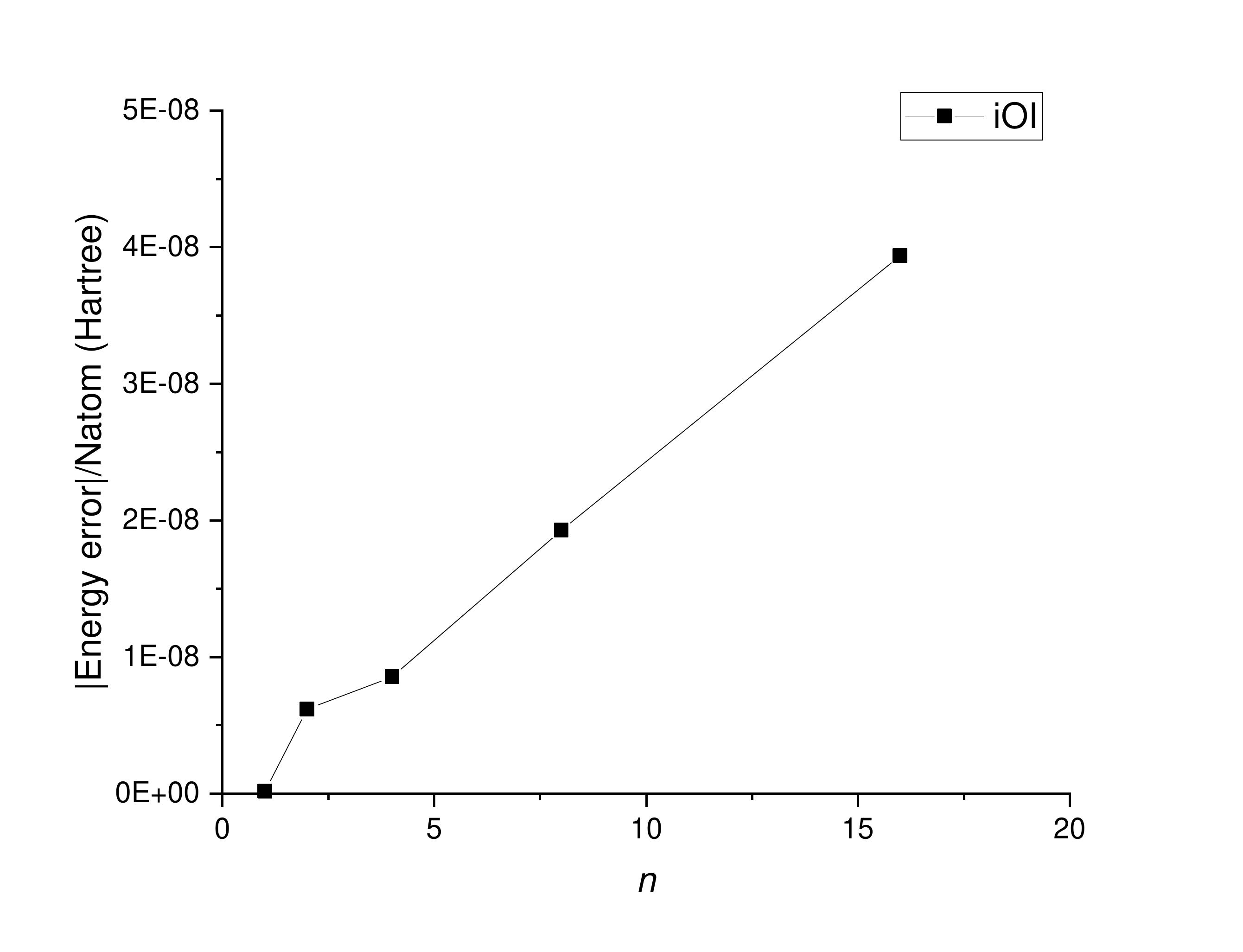}
\includegraphics[width=\textwidth]{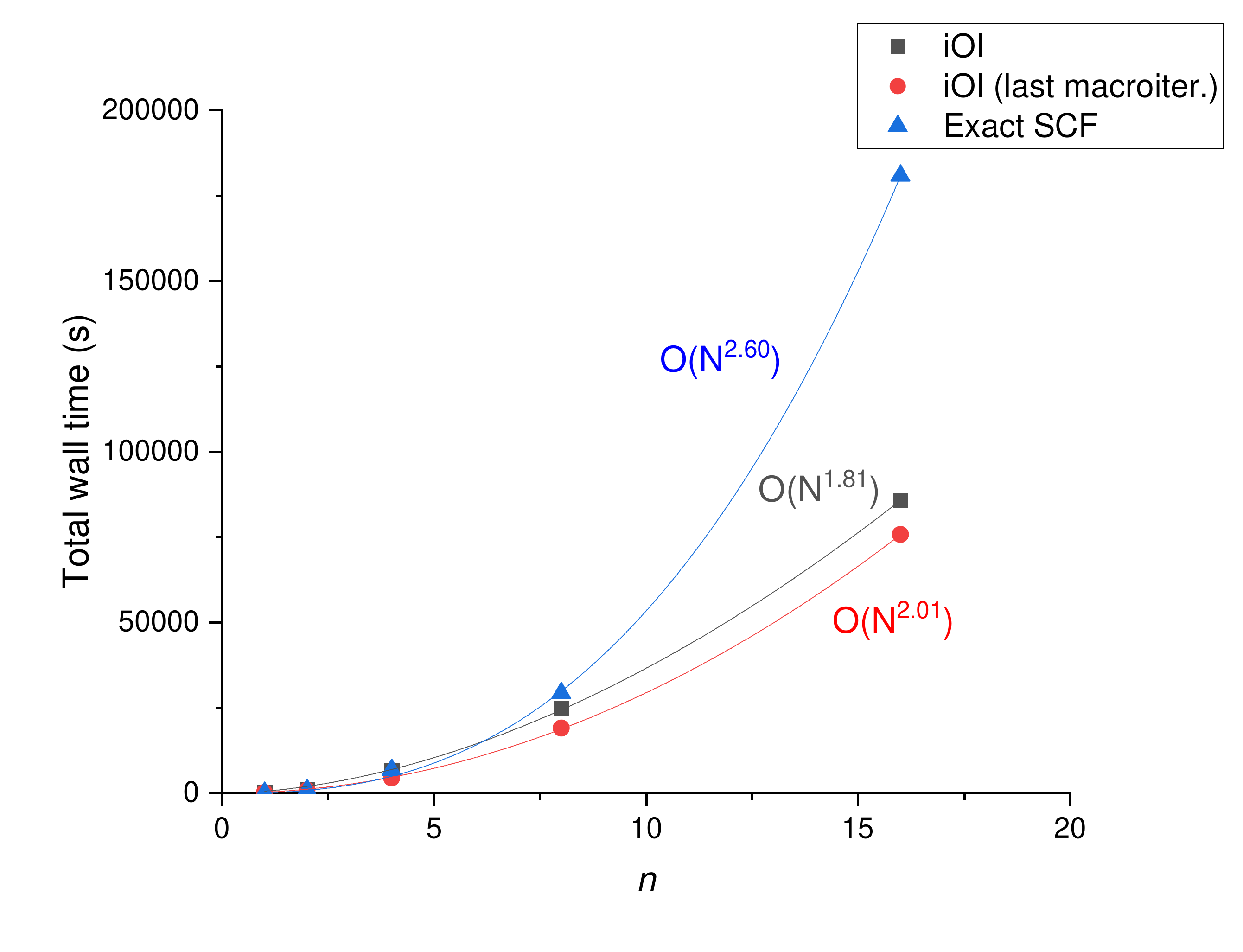}
\end{minipage}
\begin{minipage}[c]{0.45\textwidth}
\centering
\includegraphics[width=\textwidth]{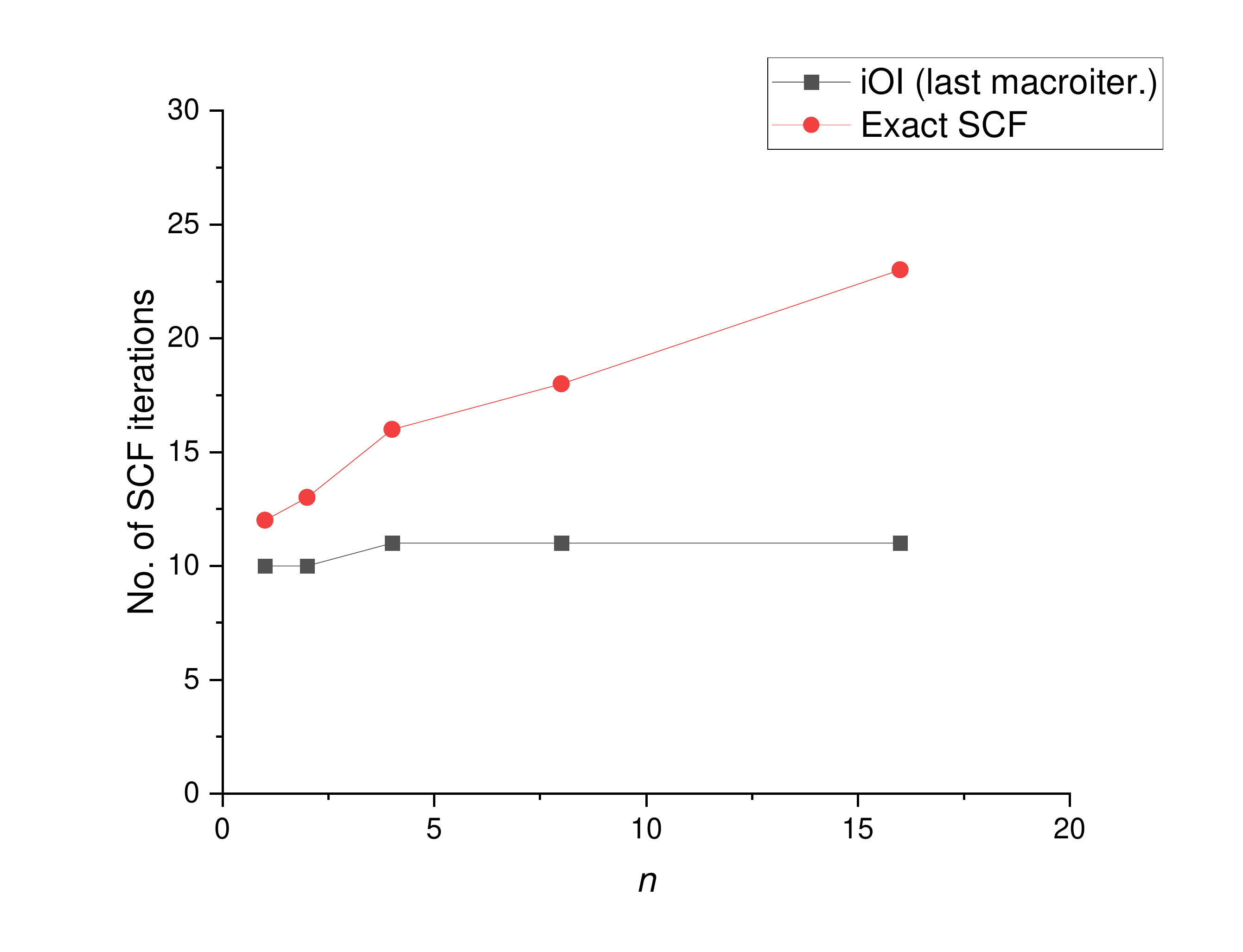}
\includegraphics[width=\textwidth]{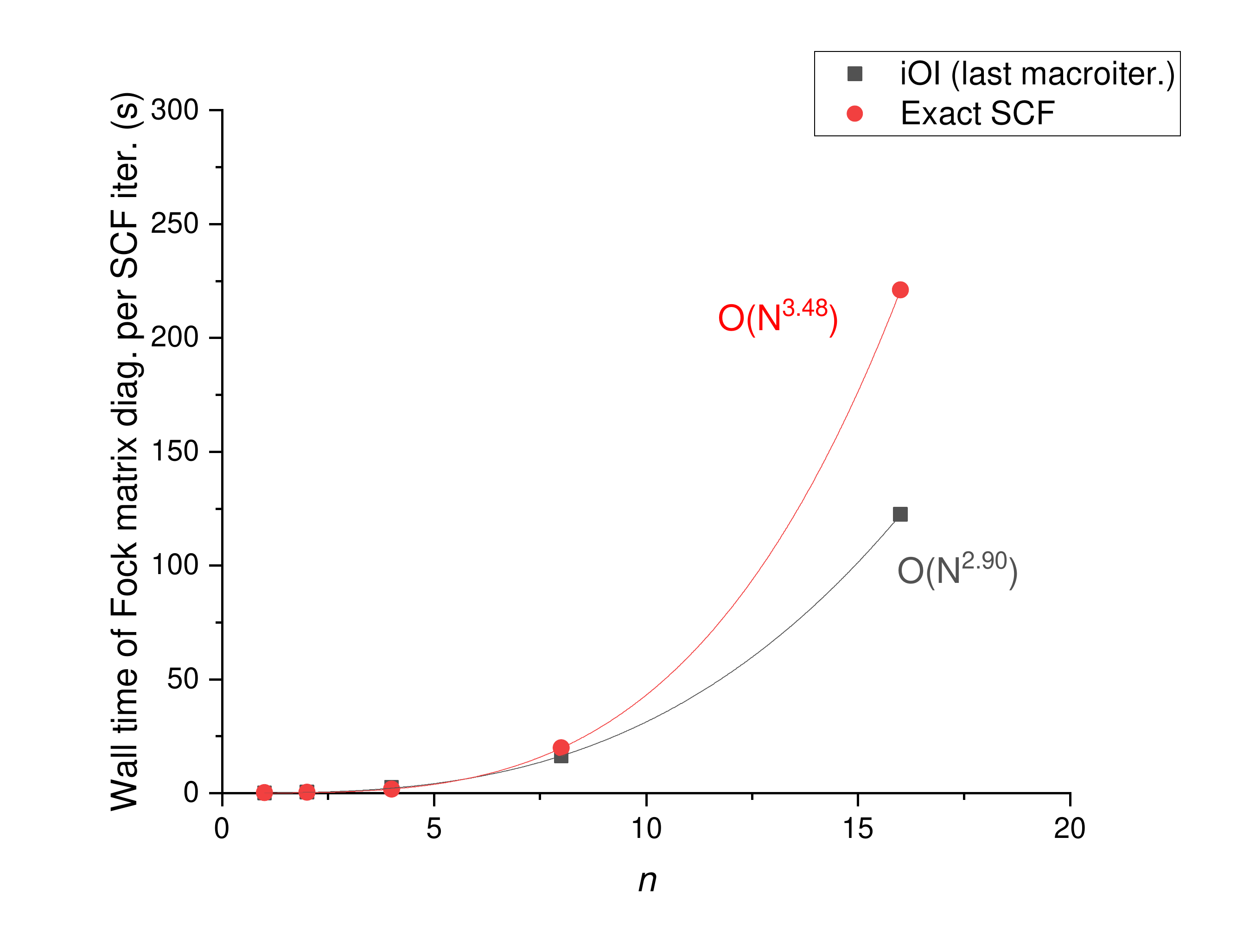}
\end{minipage}
\caption{Comparison of iOI and exact SCF calculations of DNA$_n (n=1,2,4,8,16)$. Top left: error of the iOI energy compared with
 the exact SCF energy; top right: number of SCF iterations of the final iOI macroiteration and exact SCF;
 bottom left: total wall time; bottom right: wall time of the Fock matrix (block-)diagonalization averaged over number of SCF iterations.}
\label{DNAn}
\end{figure}

To estimate the scaling behaviors of the energy error and computational time with respect to system size, we performed iOI and exact SCF calculations on the series of DNA$_n (n=1,2,4,8,16)$,
with the results summarized in Fig. \ref{DNAn}. The error of the iOI energy per atom is very small for all systems studied
but tends to increase linearly with respect to system size (Fig. \ref{DNAn}, top left).
The latter can be understood as follows: if every LMO is converged to an error threshold $\epsilon$, the (long-range) Coulomb potential felt by a given LMO will be in error approximately of  $O(\epsilon N_{\mathrm{LMO}})$, thereby leading to a linear dependence of the energy error per atom on the system size.
Intriguingly, the number of the exact SCF iterations from the atomic guess increases essentially linearly with respect to the system size,
while the number of iterations of the final iOI macroiteration is almost independent of the system size (Fig. \ref{DNAn}, top right).
This implies simply that the atomic guess becomes worse and worse as the system enlarges but the fragmental guess inherent in the iOI procedure has a constant quality.
As a result, iOI has a lower scaling than the exact SCF: while iOI is slower than the exact SCF for DNA$_1$ (62 atoms, 544 basis functions)
and DNA$_2$ (128 atoms, 1176 basis functions), a crossover is observed at DNA$_4$ (260 atoms, 2440 basis functions), and
iOI becomes about twice as fast as the exact SCF for DNA$_{16}$ (1052 atoms, 10024 basis functions) (Fig. \ref{DNAn}, bottom left). The results highlight the frequently
overlooked fact that, for an iterative method, the scaling of a single iteration does not necessarily reflect the overall scaling of the method itself, as the number of iterations can depend on the system size.
In this sense, iOI improves the overall scaling of SCF by a much reduced number of iterations, even though it does not improve the scaling of individual SCF iterations.
Moreover, the block-diagonalization is noticeably faster than the full diagonalization of the Fock matrix for the largest systems (Fig. \ref{DNAn}, bottom right), which contributes further (albeit marginally) to the speedup of iOI over the exact SCF. It deserves to be mentioned that the cubic scaling of the iOI Fock matrix block-diagonalization
is the result of using dense matrix algebra. As the matrices $\mathbf{F}$, $\mathbf{D}$ and $\mathbf{P}$ in Eqs. (\ref{F:CMO}), (\ref{DecMat}), and (\ref{NormMat}), respectively, are all numerically sparse, with only $O(N_{\mathrm{bas}})$ non-negligible elements,
the Fock matrix block-diagonalization can be made linear scaling when sparse matrix algebra is used. This is to be contrasted with the full diagonalization, which can at best be made quadratic scaling even if the sparsity of $\mathbf{F}$ is explored: the result of the full diagonalization, the CMO coefficient matrix, is a full matrix with generally $O(N_{\mathrm{bas}}^2)$ non-negligible elements, which necessarily requires at least $O(N_{\mathrm{bas}}^2)$ time to generate.

\begin{figure}
\centering
\includegraphics[width=0.7\textwidth]{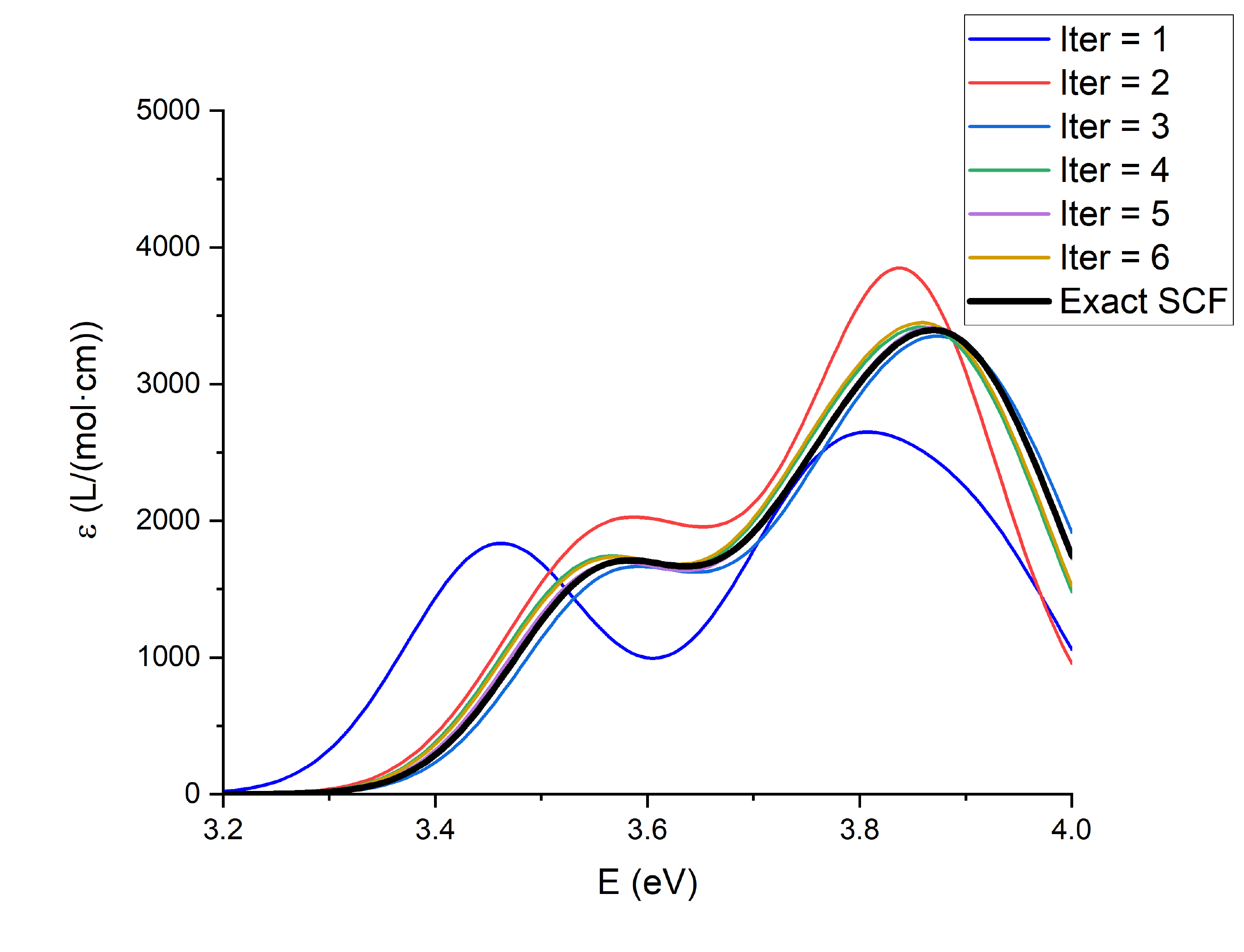}
\caption{Electronic absorption spectra of DNA$_4$ using LMOs from partially converged iOI calculations and fully converged exact SCF calculation.
All transitions are Gaussian broadened with 0.2 eV for the full width at half maximum. The numbers of SCF iterations
in the final iOI macroiteration are given in the legend. }
\label{dna4-tddft}
\end{figure}

Finally, we examine the quality of the iOI wave function (LMOs) by computing the lowest 10 excited states of DNA$_4$ with time-dependent density functional theory.
As shown in Fig. \ref{dna4-tddft}, the final iOI macroiteration, which is the most expensive step of an iOI calculation for sufficiently large molecules,
need not be fully converged: 3 SCF iterations are already enough to generate sufficiently good LMOs, in terms of which
the absorption spectrum is practically indistinguishable from that derived from the exact LMOs.
Stopped at this point, the iOI energy (-9989.825528 a.u.) is 112 mHartree (0.43 mHartree/atom) higher than the exact SCF energy (-9989.937405 a.u.);
the total elapsed time is 3566 s, about half that of the fully converged iOI (6636 s) or the exact SCF calculation (6769 s).
It can therefore be concluded that iOI not only provides virtually exact SCF energies but also excellent SCF wave functions (LMOs).

\section{Conclusions and outlook}\label{Conclusion}
A bottom-up, fragment-based SCF solver, iOI, has been developed. Unlike most fragment-based methods which conquer only the energy approximately
but not the wave function at all, iOI has the following merits:
(1) the starting small fragments are merged automatically to become just large enough for the description of the LMOs of the whole system; (2)
the number of global SCF iterations is essentially independent of system size, in contrast to
the traditional SCF approach which requires more iterations for large than for small systems;
(3) it can yield not only the exact SCF energy with notable time savings but also fully converged occupied and virtual LMOs without additional cost.
Some additional merits of iOI, which will be demonstrated in future, include (a) facilitation of SCF convergence in difficult situations, (b)
directed convergence to a user-specified configuration when
there exist multiple SCF solutions or when an excited configuration is desired, and (c) automatic determination of
the size of active space in multi-scale calculations of exceedingly large systems.


\section*{Acknowledgement}
This work was supported by the National Key R\&D Program of
China (Grant No. 2017YFB0203402), National Natural Science Foundation of China (Grant Nos. 21833001 and 21973054),
Mountain Tai Climb Program of Shandong Province, and Key-Area Research and Development Program of Guangdong Province (Grant No. 2020B0101350001).

\section*{Data Availability Statement}
The data that supports the findings of this study are available within the article.

\appendix

\section{Automatic selection of buffer atoms for each fragment} \label{autofrag}
\begin{figure}
\centering
\includegraphics[width=\textwidth]{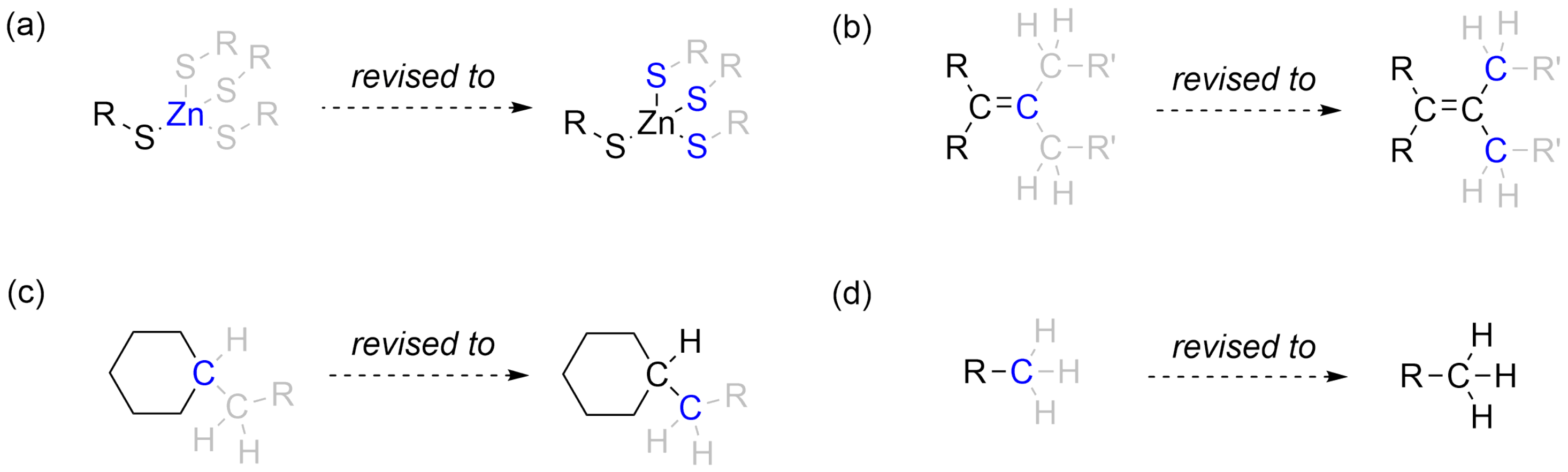}
\caption{Special treatment of dangling bonds: (a) the PHO boundary atom is not a $p$-block element; (b) the dangling bond is not a single bond; (c) the PHO boundary atom is connected to more than one buffer atoms; (d) all environment atoms bonded to a PHO boundary atom are hydrogens. The central and buffer atoms are indicated in black, while the PHO boundary atoms are indicated in blue.}
\label{buffering_special_cases}
\end{figure}

In Steps (\ref{buffering}) and (\ref{add_pho}) of the iOI algorithm, some atoms that are sufficiently close to a pFrag are joined with the pFrag to form a subsystem, so as to provide a more realistic environment for the pFrag on one hand and to saturate the dangling bonds on the other. The term ``sufficiently close'' can however adopt many possible definitions. The most straightforward way is to define ``close'' as being spatially close. Thus, for each atom in the pFrag, one can draw a sphere of a certain radius (hereafter called the buffer radius), $R_L^{(\tilde{m})}$ (where $\tilde{m}$, the macroiteration number, is equal to 0 here), centered at that atom. Any atom external to the fragment that falls into at least one of the spheres
\begin{equation}
A\in \mathcal{B}_L^{(\tilde{m})} \quad\mathrm{iff}\quad \exists B\in \mathcal{F}_L^{(\tilde{m})} \quad\mathrm{s.t.}\quad R_{AB} < R_L^{(\tilde{m})} \label{real_space_metric}
\end{equation}
is then chosen as a buffer atom.
The buffering process is then followed by saturating the remaining dangling bonds by PHO boundary atoms to give the cap $\mathcal{C}_L^{(\tilde{m})}$, which is the union of the buffer atoms and the PHO boundary atoms. This approach however has a few drawbacks:
\begin{enumerate}[(1)]
\item The importance of an atom as a buffer atom is not always correlated with its spatial distance from the pFrag. Rather, orbital-based metrics are usually more reliable than real space-based metrics. Since one of the purposes of introducing the buffer atoms is to provide enough AOs for accurately expanding the pFLMOs, the most rigorous metric would be the population of the pFLMOs on the AOs of the buffer atoms, i.e., a buffer atom is only included if the pFLMOs have a large enough population on that atom. However, the pFLMOs are not known at the time the buffer atoms are chosen. Thus, we instead measure the importance of a buffer atom by the overlap matrix elements of the AOs of the buffer atom with the AOs of the pFrag. Specifically, we define the following ``effective distance'' between two atoms: 
\begin{equation}
R_{AB}^{\mathrm{eff}} = \kappa\sqrt{-\ln \max_{\mu\in A, \nu\in B} |S_{\mu\nu}|}, \quad \kappa = 2.0 \mbox{ \AA}. \label{smetric}
\end{equation}
Herein $\kappa$ is chosen such that $R_{AB}^{\mathrm{eff}}$ is similar to the real-space distance between the atoms $A$ and $B$ (in Angstrom) when typical double-zeta basis sets are used. We then choose the buffer atoms via Eq. \eqref{real_space_metric}, but using $R_{AB}^{\mathrm{eff}}$ in Eq. \eqref{smetric} in place of $R_{AB}$. The threshold radius at macroiteration 0, $R_L^{(0)}$, is chosen to be 2.0~\AA irrespective of the subsystem $L$. However, $R_L^{(\tilde{m})}$ will become subsystem-dependent when $\tilde{m}>0$ (vide infra).

\item In case that the use of PHO boundary atoms to saturate the dangling bonds is inappropriate,
the corresponding PHO boundary atoms are redesignated as buffer atoms, and the search for dangling bonds is performed again. This happens under four circumstances:
\begin{enumerate}
\item The PHO boundary atom is not a $p$-block element atom (see Fig. \ref{buffering_special_cases}(a)). This is because the PHO algorithm involves the hybridization of the valence $s$ and $p$ orbitals to yield 4 hybrid orbitals, which makes sense only if the PHO boundary atoms are $p$-block element atoms.

\item The dangling bond is a multiple bond (see Fig. \ref{buffering_special_cases}(b)). This is because the PHO algorithm works only for the $\sigma$ type of active hybrid orbitals.

\item The same PHO boundary atom is used to saturate more than one dangling bond (Fig. \ref{buffering_special_cases}(c)). While the original PHO method\cite{PHO} supports PHO boundary atoms with more than one active hybrid orbitals, the present implementation is confined to PHO boundary atoms with a single active hybrid orbital.

\item All the environment atoms of a PHO boundary atom are hydrogen atoms (Fig. \ref{buffering_special_cases}(d)). This usually happens when the PHO boundary atom is a methyl carbon atom. In this case, it is a better strategy to convert the carbon atom to a buffer atom and include the three methyl hydrogen atoms in the buffer, since this hardly increases the computational cost but makes the description of the methyl group much more reliable. Note that this includes PHO boundary atoms that have zero environment atoms (e.g. halogens) as a special case.
\end{enumerate}
\end{enumerate}

\section{Modified PHO} \label{pho}

The PHO method developed by Gao et al.\cite{PHO} is a convenient way for saturating dangling bonds in QM/MM calculations. Compared with the use of link atoms, the PHO method suffers less from artifacts such as over-polarization of the QM region by the MM region. Herein we introduce some modifications to the PHO approach to
describe more complex chemical environments.

\begin{figure}
\centering
\includegraphics[width=0.8\textwidth]{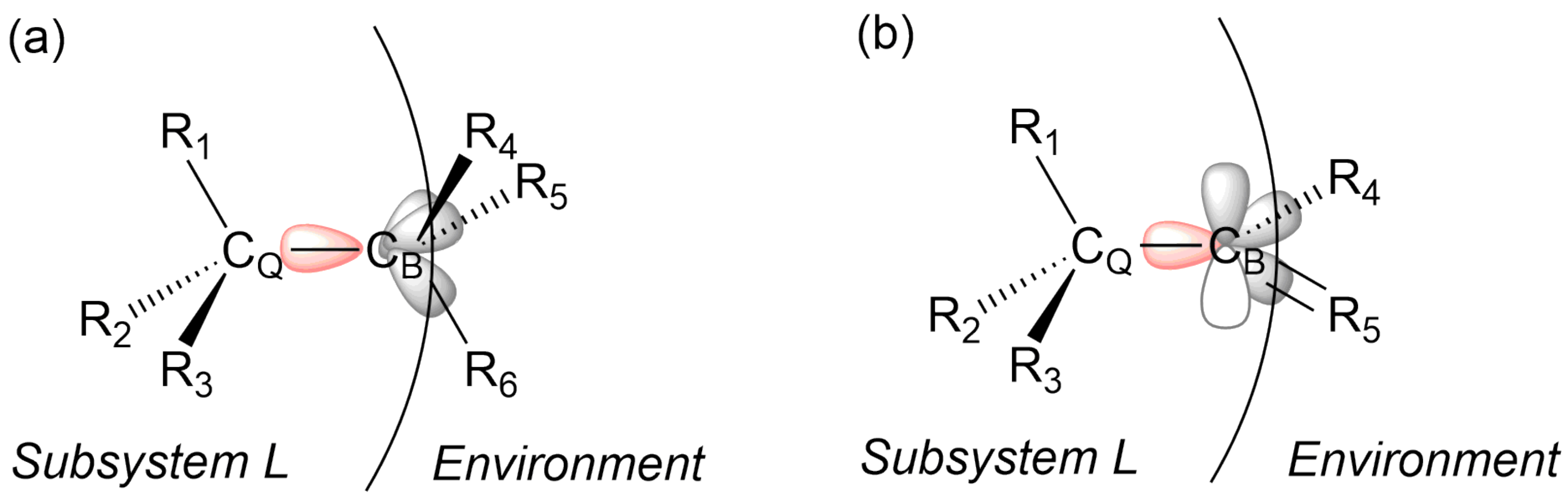}
\caption{The boundary of two subsystems capped by PHO boundary atoms (C$_{\mathrm{B}}$): (a) the dangling bond is C(sp$^3$)-C(sp$^3$); (b) the dangling bond is C(sp$^3$)-C(sp$^2$). The active and auxiliary hybrid orbitals are colored in red and grey, respectively.}
\label{pho_example}
\end{figure}

Suppose that the global system is fragmented at a C-C bond (Fig. \ref{pho_example}). Traditionally,
the free valence of the C$_{\mathrm{Q}}$ atom is saturated by a hydrogen atom; the atoms on the left side of
the bond are thus assigned to subsystem $L$, while those on the right side are termed
environment. In the PHO method, however, atom C$_{\mathrm{B}}$ is included in subsystem $L$ and is called PHO boundary atom. The four valence orbitals (one $2s$ and three $2p$ orbitals) of C$_{\mathrm{B}}$ are hybridized to form four hybrid orbitals (four $sp^3$ orbitals in Fig. \ref{pho_example}(a); three $sp^2$ orbitals and one $p$ orbital in Fig. \ref{pho_example}(b)). The one that is oriented towards C$_{\mathrm{Q}}$ is termed active, whereas the other three are termed auxiliary. The former is allowed to mix with the AOs of the remaining parts of the subsystem, whereas the latter represent free valences of C$_{\mathrm{B}}$.
Since they are actually saturated in the whole system by bonding with the \ce{Rx} groups, the auxiliary hybrid orbitals must be frozen, i.e.,
projected out of the variational space. Otherwise one is effectively calculating a carbon triradical \ce{R1R2R3C-C}$\vdots$, instead of the closed-shell system \ce{R1R2R3C-CR4R5R6} or \ce{R1R2R3C-CR4R5}. Nevertheless, the auxiliary hybrid orbitals can accommodate electrons, so as to
provide an effective external potential for the rest of the subsystem.

To uniquely define our modified PHO method, several details need to be addressed:
\begin{enumerate}[(1)]
\item Generation of valence atomic orbitals\\
 While in minimal basis set calculations some AOs can be unambiguously assigned as the atomic valence orbitals,
 this is generally not true for extended basis sets. In the original PHO method\cite{PHO}, the STO-3G $2s$ and $2p$
 are taken to be the atomic valence orbitals, even when the calculation is performed with extended basis sets. Although the deficiency of the STO-3G basis
 is partly remedied by transforming the density matrix elements of the PHO boundary atom from STO-3G to the actually employed basis,
 this algorithm can still be potentially unreliable when the STO-3G orbitals are quite different from the exact
 atomic valence orbitals (e.g., when the PHO boundary atom is a heavy element). To resolve this issue,
 we perform a spherically averaged and unpolarized atomic ROHF/ROKS calculation for each PHO boundary atom with the actual basis, so that the valence orbitals of the atomic calculation form a minimal valence basis which is of much better quality than STO-3G. We choose only $p$-block atoms as PHO boundary atoms.

\item Generation of hybrid orbitals\\
 Hybridizing the valence $s$ and $p$ orbitals is trivial if the boundary atom adopts a perfect tetrahedral ($sp^3$), trigonal ($sp^2$), or linear ($sp$) geometry. However, real-world molecules seldom possess such exact symmetries. In particular, if the PHO boundary atom has a lone pair,
 the deviations from exact symmetries can be quite significant.
 Before discussing the generation of hybrid orbitals under nonideal symmetries, we first introduce the concept of a \emph{hybrid orbital pointing towards} an atom.
 A hybrid orbital of atom $B$ of the following form,
\begin{equation} \label{HO}
\psi^B = c_s^B\chi_s^B + c_{p_x}^B\chi_{p_x}^B + c_{p_y}^B\chi_{p_y}^B + c_{p_z}^B\chi_{p_z}^B,
\end{equation}
is said to be pointing towards atom $A$, if its coefficients satisfy $(c_{p_x}^B,c_{p_y}^B,c_{p_z}^B) = k\vec{r}_{BA}$, with $\vec{r}_{BA}$ being
 the vector from atom B to atom A and $k$ being a positive constant of proportionality.
 In order for the definition to make sense, the phases of the atomic valence orbials $\{\chi_s^B,\chi_{p_x}^B,\chi_{p_y}^B,\chi_{p_z}^B\}$ must be such that when a hybrid orbital of atom $B$ is pointing towards atom $A$, the lobe of that hybrid orbital to the direction of $A$ is larger than the lobe pointing away from $A$. Intuitively, when a hybrid orbital of atom $B$ points towards atom $A$, its $\sigma$-overlap with the AOs of atom $A$ is maximized,
 which is the most energetically favorable scenario for bonding. Thus, for a $p$-block element atom $B$ bonded to $n$ atoms,
 we expect that $n$ hybrid orbitals of the atom will point (at least approximately) towards the atoms directly bonded to $B$, and
 the remaining $4-n$ hybrid orbitals are either non-bonding orbitals or form $\pi$ bonds with the neighboring atoms.
 While for $n \leq 3$ the former hybrid orbitals can indeed be made exactly pointing towards the neighboring atoms, for $n=4$ this
 is generally not possible because the problem is overdetermined: requiring all four hybrid orbitals to point towards specified
 atoms introduces 8 constraints on the AO coefficients of the hybrid orbitals (2 for each hybrid orbital), while
 the orthonormality conditions of the four hybrid orbitals introduce another $4\times(4+1)/2=10$ constraints,
 giving a total of 18 constraints, but there are only 16 variables to be determined (4 for each hybrid orbital).
 This is to be contrasted with the $n=3$ case, where there are exactly 16 constraints, giving a well-determined problem.
 The $n=2$ case is accordingly underdetermined, which can also be problematic as the solution to the hybrid orbitals is not unique.

Gao's approach\cite{PHO,GHO} for $n=4$ amounts to requiring three of the hybrid orbitals to point towards the three environment atoms \ce{R4}, \ce{R5}, and \ce{R6},
so that the fourth hybrid orbital does not necessarily point exactly towards the QM atom C$_{\mathrm{Q}}$ (Fig. \ref{pho_example}).
Thus, one of the ``exactly pointing'' constraints is relaxed, making the problem well-determined.
It was found by Exner and coworkers\cite{SPX} that this will deteriorate the description of the C$_{\mathrm{Q}}$-C$_{\mathrm{B}}$ bond,
especially if the local geometry of the boundary atom C$_{\mathrm{B}}$ is significantly distorted from ideal tetrahedral geometry.
Therefore, they proposed to force three of the hybrid orbitals to point towards C$_{\mathrm{Q}}$, \ce{R4} and \ce{R5},
 at the expense that the fourth hybrid orbital will not point exactly towards \ce{R6}.
 This approach improves the accuracy for some molecules, but the price to pay is that the result is no
 longer invariant with respect to permutations of the environment atoms \ce{R4}, \ce{R5}, and \ce{R6},
 which is clearly undesirable. They then proposed to fix the latter problem by generating
 a set of hybrid orbitals for each of these permutations, and average the resulting hybrid orbitals,
 followed by a symmetric orthogonalization, which solves the aforementioned problem at the expense of increased complexity of the method.
Herein, we propose a simple and elegant remedy to this problem by generating four hybrid orbitals $\{\tilde{\psi}^B\}$ that are normalized but not necessarily mutually orthogonal:
\begin{equation}
\tilde{\psi}^B = \tilde{c}_s^B\chi_s^B + \tilde{c}_{p_x}^B\chi_{p_x}^B + \tilde{c}_{p_y}^B\chi_{p_y}^B + \tilde{c}_{p_z}^B\chi_{p_z}^B, \nonumber
\end{equation}
\begin{equation}
\tilde{c}_s^2 = 1/4, \quad \tilde{c}_{p_x}^2 + \tilde{c}_{p_y}^2 + \tilde{c}_{p_z}^2 = 3/4.\label{mPHO}
\end{equation}
To be unique, $\{\tilde{\psi}^B\}$ are required to be $sp^3$-hybridized, as implied by the latter conditions in Eq. \eqref{mPHO}. Being nonorthogonal,
$\{\tilde{\psi}^B\}$ can be made to point exactly towards the respective atoms. They are then symmetrically orthogonalized to yield the final
 orthonormal hybrid orbitals. Since the symmetric orthogonalization procedure produces orbitals that are closest to the original ones,
the present hybrid orbitals are more realistic than the previous ones\cite{PHO,GHO,SPX},
especially for heavily distorted structures (which will be demonstrated in a separate paper).
%

When $n=3$ or $n=2$, we still generate the $n$ hybrid orbitals $\{\psi^B\}$ by the
above procedure (i.e., the initial hybrid orbitals $\{\tilde{\psi}^B\}$ are still taken as $sp^3$-hybridized),
and find the remaining $4-n$ hybrid orbitals by Schmidt orthogonalization of the atomic valence $s$ and $p$
orbitals against the $n$ hybrid orbitals $\{\psi^B\}$. Note that this works even when the real hybrid orbitals
are not $sp^3$-hybridized, because the symmetric orthogonalization step automatically adjusts the $s$ components of
the hybrid orbitals. For example, for a trigonal planar PHO boundary atom like the C$_{\mathrm{B}}$ atom in Fig. \ref{pho_example}(b),
we generate three nonorthonormal $sp^3$ hybrid orbitals $\{\tilde{\psi}^B\}$ that point towards the atoms C$_{\mathrm{Q}}$, \ce{R4},
and \ce{R5}, respectively. If the geometry of C$_{\mathrm{B}}$ is exactly threefold symmetric,
each hybrid orbital must have exactly $1/3$ of $s$ character to be mutually orthogonal, thereby giving the desired three $sp^2$ hybrid orbitals.
The Schmidt orthogonalization step then finds the $p$ orbital perpendicular to the trigonal plane as the
fourth hybrid orbital, since this is the orthogonal complement of the aforementioned three $sp^2$ hybrid orbitals
in the four-dimensional space of the atomic valence orbitals of C$_{\mathrm{B}}$.

\item Projection of auxiliary hybrid orbitals out of variational space\\
This is required under two circumstances, one is the projection of the components of the auxiliary hybrid orbitals from the pFLMOs
before the elimination of linear dependence, to ensure that the linearly independent pFLMOs are free of contributions
from the auxiliary hybrid orbitals, and the other is the projection of the contributions of the auxiliary hybrid orbitals from
the Fock matrix before (block-)diagonalization (Steps (\ref{SCF_macroiter0}) and (\ref{SCF_macroiterN})),
to ensure that the new canonical (for macroiteration 0) or localized (for macroiteration $\tilde{m}>0$) MOs from
the (block-)diagonalization of the Fock matrix are free of contributions from the auxiliary hybrid orbitals.
The projection is trivial when there is only one PHO boundary atom in the system, where the auxiliary hybrid orbitals
are mutually orthogonal, as described by Gao\cite{PHO}. When there are more than one PHO boundary atoms,
the auxiliary hybrid orbitals that belong to different PHO boundary atoms are generally not exactly orthogonal
with each other (although usually very nearly so, since the PHO boundary atoms of a subsystem are usually relatively far from each other).
In this case, they must be first symmetrically orthogonalized to give the orthonormal auxiliary hybrid orbitals $\{\psi_{\mu}^{B,\mathrm{PHO}}\}$.
Then, we project out the contributions of $\{\psi_{\mu}^{B,\mathrm{PHO}}\}$ from the pFLMOs and/or the Fock matrix, as usually done\cite{TennoGHO}.
\item Occupation assignment of auxiliary hybrid orbitals\\
The original PHO method\cite{PHO}, which was developed in the context of QM/MM,
requires that the atomic charge of the PHO boundary atom $q_B$ be known before the SCF iterations,
which is generally available from the definition of the MM force field. Then, if $n=4$ (i.e., the number of auxiliary hybrid orbitals is 3),
the occupation numbers are determined as
\begin{equation} 
f(\psi_{\mu}^{B,\mathrm{PHO}}) = 1 - q_B/3, \label{pho_occupation}
\end{equation}
so that if the bond C$_{\mathrm{Q}}$-C$_{\mathrm{B}}$ is exactly non-polar,
the QM charge of the atom C$_{\mathrm{B}}$ reproduces the MM charge $q_B$. The description of the charge of
the PHO boundary atom by fractional occupations of the auxiliary hybrid orbitals, rather than by point charges,
is the key to avoid over-polarization effects at the boundary of the QM region\cite{PHO}. However,
in macroiteration 0 of the iOI algorithm, the charge $q_B$ is not known in advance and is set simply to zero.
For macroiteration $\tilde{m}>0$, we identify the subsystem at macroiteration $\tilde{m}-1$ whose pFrag
contains the PHO boundary atom in question. The L\"owdin charge of C$_{\mathrm{B}}$ in that subsystem
is then assigned to $q_B$.

For $n=3$ or $n=2$, it is generally advocated to partition the occupation numbers unevenly between
the hybrid orbitals when some of the auxiliary hybrid orbitals are lone pairs, e.g.,
the occupation numbers are partitioned in a 2:1:1 ratio among the three auxiliary hybrid orbitals for amine nitrogen atom,
with the first hybrid orbital being the lone pair orbital.
The ratio can be even more delicately tuned by considering the electronegativities of the neighboring atoms\cite{SPX,TennoGHO}.
However there are many situations where the ratio of occupation numbers among the auxiliary hybrid orbitals
cannot be unambiguously assigned, such as when the PHO boundary atom is an amide nitrogen atom,
where the occupation of the non-bonding auxiliary hybrid orbital is difficult to assign
accurately without inventing a rule specifically for amides. Therefore, for simplicity,
we always partition the fractional occupations of a PHO boundary orbital
 evenly across all auxiliary hybrid orbitals regardless of the nature of the PHO boundary atom.
 As will be demonstrated in a separate paper, the results are not noticeably inferior to
 those schemes where the occupation numbers are partitioned unevenly, at least when only
 the quality of the LMOs are concerned.
\item Coulomb interaction between the subsystem and environment\\
 In the original PHO implementation\cite{PHO} one generally describes all MM atoms as point charges, which can incur a non-trivial cost when the MM region is very large. In the iOI algorithm, however, it is only necessary to accurately describe the polarization effect of the point charges on the subsystem, but not the electrostatic energy between the point charges and the subsystem, since the electronic energy is ultimately obtained by a global SCF calculation. As the polarization effect is much more short-ranged ($R^{-4}$) than the electrostatic energy ($R^{-1}$), we find it only necessary to include the point charges of the environment atoms that are directly bonded to the PHO boundary atoms. As before, the point charges are zero for macroiteration 0 and are taken as the L\"owdin charges from the previous macroiteration for macroiteration $\tilde{m}>0$.

\end{enumerate}

\section{Merging subsystems} \label{merging}

As described in Step (\ref{ioimerge}) of the iOI algorithm, in the end of each macroiteration, the pFrags will be merged together to give new pFrags, in a way that
every new pFrag consists of either one or two old pFrags, which is followed by a capping step. The merging process only makes sense if the two old pFrags
comprising a new one are spatially adjacent. This can be a non-trivial condition to fulfill for molecules with complex topologies.
Consider, e.g., a star-shaped molecule where a pFrag $\mathcal{F}_L$ is connected to three mutually disconnected pFrags $\mathcal{F}_{M_1}$, $\mathcal{F}_{M_2}$ and $\mathcal{F}_{M_3}$. If the three pFrags are sufficiently similar but not identical, then it is not easy to decide which of them
should be merged with $\mathcal{F}_L$. Choosing the spatially closest one to merge with $\mathcal{F}_L$ is not always the best solution for
the remaining two pFrags may be too far from each other. Keeping such situations in mind, we propose the following merging process:
\begin{enumerate}[(1)]
\item Make a list of the pFrags of all not yet converged subsystems, the ordering of which is immaterial;

\item If the number of nonconvergent subsystems is odd, there will be at least one pFrag that cannot be merged with others. To make the sizes of the new pFrags as uniform as possible, the pFrag with the largest number of basis functions is removed from the list, such that it is directly carried over to the next macroiteration as a new pFrag. The remaining pFrags can then be merged in a pairwise fashion;

\item Calculate the distance $d(\mathcal{F}_L^{(\tilde{m})},\mathcal{F}_M^{(\tilde{m})})$ between ever pair of pFrags, $\mathcal{F}_L^{(\tilde{m})}$ and $\mathcal{F}_M^{(\tilde{m})}$, which is defined as the minimum
 of the effective distances, Eq. \eqref{smetric}, between atoms in $\mathcal{F}_L^{(\tilde{m})}$ and in $\mathcal{F}_M^{(\tilde{m})}$;
\item Starting from the first pFrag in the list, form a pFrag pair with pFrag $\mathcal{F}_{L'}^{(\tilde{m})}$ and its spatially closest pFrag $\mathcal{F}_L^{(\tilde{m})}$.
Remove $\mathcal{F}_L^{(\tilde{m})}$ and $\mathcal{F}_{L'}^{(\tilde{m})}$ from the list and repeat the process for the next surviving entry of the list;

\item For each \emph{pair of pFrag pairs} $\{\{\mathcal{F}_L^{(\tilde{m})},\mathcal{F}_{L'}^{(\tilde{m})}\},\{\mathcal{F}_M^{(\tilde{m})},\mathcal{F}_{M'}^{(\tilde{m})}\}\}$, permute the four pFrags so as to reduce the maximum intrapair distance $\max(d(\mathcal{F}_L^{(\tilde{m})},\mathcal{F}_{L'}^{(\tilde{m})}),d(\mathcal{F}_M^{(\tilde{m})},\mathcal{F}_{M'}^{(\tilde{m})}))$.
    This is repeated for every possible pair of pFrag pairs, until the maximum intrapair distance cannot be reduced by permutations. There is however an exception: if the minimum value for the maximum intrapair distance is larger than a threshold, $d_{\mathrm{max}} = 4.0$~\AA, no more than one of the two pFrag pairs will actually give rise to a new pFrag (see next step). In this case we minimize $\min(d(\mathcal{F}_L^{(\tilde{m})},\mathcal{F}_{L'}^{(\tilde{m})}),d(\mathcal{F}_M^{(\tilde{m})},\mathcal{F}_{M'}^{(\tilde{m})}))$ instead of $\max(d(\mathcal{F}_L^{(\tilde{m})},\mathcal{F}_{L'}^{(\tilde{m})}),d(\mathcal{F}_M^{(\tilde{m})},\mathcal{F}_{M'}^{(\tilde{m})}))$. The pFrag pairs are then as mutually proximal as possible. Another benefit of this step is that it minimizes (though not necessarily eliminates) the dependence of the result on the ordering of the pFrags in the list;
\item For each pFrag pair $\{\mathcal{F}_L^{(\tilde{m})},\mathcal{F}_M^{(\tilde{m})}\}$, if $d(\mathcal{F}_L^{(\tilde{m})},\mathcal{F}_M^{(\tilde{m})})\leq d_{\mathrm{max}}$, merge the two pFrags to form a new pFrag. If $d(\mathcal{F}_L^{(\tilde{m})},\mathcal{F}_M^{(\tilde{m})}) > d_{\mathrm{max}}$, the two pFrags are not considered to be adjacent and
     are therefore carried over directly to the next macroiteration as two new pFrags;
\item Generate the caps $\mathcal{C}_L^{(\tilde{m}+1)}$ for all the new pFrags $\mathcal{F}_L^{(\tilde{m}+1)}$, according to Eq. \eqref{real_space_metric}. The buffer radii, $R_L^{(\tilde{m}+1)}$, are calculated as follows: (a) set $R_L^{(\tilde{m}+1)} = \max_I R_{\mathcal{P}_L^{(\tilde{m}+1)}(I)}^{(\tilde{m})}$, i.e., the maximum of the buffer radii of the parental subsystems of $\mathcal{S}_L^{(\tilde{m}+1)}$; (b) increase $R_L^{(\tilde{m}+1)}$ until exactly one additional atom is included in the buffer; (c) increase $R_L^{(\tilde{m}+1)}$ by 1.0~\AA. Step (b) guarantees that the new subsystem $\mathcal{S}_L^{(\tilde{m}+1)}$ is strictly larger than its parental subsystems, while Step (c) ensures that the incremental cap $\Delta\mathcal{C}_L^{(\tilde{m}+1)}$ is not too small.
\end{enumerate}

\clearpage
\newpage

\bibliography{BDFlib}

\providecommand{\latin}[1]{#1}
\makeatletter
\providecommand{\doi}
  {\begingroup\let\do\@makeother\dospecials
  \catcode`\{=1 \catcode`\}=2 \doi@aux}
\providecommand{\doi@aux}[1]{\endgroup\texttt{#1}}
\makeatother
\providecommand*\mcitethebibliography{\thebibliography}
\csname @ifundefined\endcsname{endmcitethebibliography}
  {\let\endmcitethebibliography\endthebibliography}{}
\begin{mcitethebibliography}{123}
\providecommand*\natexlab[1]{#1}
\providecommand*\mciteSetBstSublistMode[1]{}
\providecommand*\mciteSetBstMaxWidthForm[2]{}
\providecommand*\mciteBstWouldAddEndPuncttrue
  {\def\EndOfBibitem{\unskip.}}
\providecommand*\mciteBstWouldAddEndPunctfalse
  {\let\EndOfBibitem\relax}
\providecommand*\mciteSetBstMidEndSepPunct[3]{}
\providecommand*\mciteSetBstSublistLabelBeginEnd[3]{}
\providecommand*\EndOfBibitem{}
\mciteSetBstSublistMode{f}
\mciteSetBstMaxWidthForm{subitem}{(\alph{mcitesubitemcount})}
\mciteSetBstSublistLabelBeginEnd
  {\mcitemaxwidthsubitemform\space}
  {\relax}
  {\relax}

\bibitem[Li \latin{et~al.}(2014)Li, Li, Suo, and Liu]{ACR-FLMO}
Li,~Z.; Li,~H.; Suo,~B.; Liu,~W. Localization of molecular orbitals: from
  fragments to molecule. \emph{Acc. Chem. Res.} \textbf{2014}, \emph{47},
  2758--2767\relax
\mciteBstWouldAddEndPuncttrue
\mciteSetBstMidEndSepPunct{\mcitedefaultmidpunct}
{\mcitedefaultendpunct}{\mcitedefaultseppunct}\relax
\EndOfBibitem
\bibitem[H{\o}yvik and J{\o}rgensen(2016)H{\o}yvik, and
  J{\o}rgensen]{LMOChemRev}
H{\o}yvik,~I.-M.; J{\o}rgensen,~P. Characterization and generation of local
  occupied and virtual Hartree-Fock orbitals. \emph{Chem. Rev.} \textbf{2016},
  \emph{116}, 3306--3327\relax
\mciteBstWouldAddEndPuncttrue
\mciteSetBstMidEndSepPunct{\mcitedefaultmidpunct}
{\mcitedefaultendpunct}{\mcitedefaultseppunct}\relax
\EndOfBibitem
\bibitem[Prodan and Kohn(2005)Prodan, and Kohn]{Nearsightedness}
Prodan,~E.; Kohn,~W. Nearsightedness of electronic matter. \emph{Proc. Nat.
  Acad. Sci.} \textbf{2005}, \emph{102}, 11635--11638\relax
\mciteBstWouldAddEndPuncttrue
\mciteSetBstMidEndSepPunct{\mcitedefaultmidpunct}
{\mcitedefaultendpunct}{\mcitedefaultseppunct}\relax
\EndOfBibitem
\bibitem[Warshel and Karplus(1972)Warshel, and Karplus]{QMMM1}
Warshel,~A.; Karplus,~M. Calculation of ground and excited state potential
  surfaces of conjugated molecules. I. Formulation and parametrization.
  \emph{J. Am. Chem. Soc.} \textbf{1972}, \emph{94}, 5612--5625\relax
\mciteBstWouldAddEndPuncttrue
\mciteSetBstMidEndSepPunct{\mcitedefaultmidpunct}
{\mcitedefaultendpunct}{\mcitedefaultseppunct}\relax
\EndOfBibitem
\bibitem[Day \latin{et~al.}(1996)Day, Jensen, Gordon, Webb, Stevens, Krauss,
  Garmer, Basch, and Cohen]{EFP1996}
Day,~P.~N.; Jensen,~J.~H.; Gordon,~M.~S.; Webb,~S.~P.; Stevens,~W.~J.;
  Krauss,~M.; Garmer,~D.; Basch,~H.; Cohen,~D. An effective fragment method for
  modeling solvent effects in quantum mechanical calculations. \emph{J. Chem.
  Phys.} \textbf{1996}, \emph{105}, 1968--1986\relax
\mciteBstWouldAddEndPuncttrue
\mciteSetBstMidEndSepPunct{\mcitedefaultmidpunct}
{\mcitedefaultendpunct}{\mcitedefaultseppunct}\relax
\EndOfBibitem
\bibitem[Gordon \latin{et~al.}(2001)Gordon, Freitag, Bandyopadhyay, Jensen,
  Kairys, and Stevens]{EFP2001}
Gordon,~M.~S.; Freitag,~M.~A.; Bandyopadhyay,~P.; Jensen,~J.~H.; Kairys,~V.;
  Stevens,~W.~J. The effective fragment potential method: A QM-based MM
  approach to modeling environmental effects in chemistry. \emph{J. Phys. Chem.
  A} \textbf{2001}, \emph{105}, 293--307\relax
\mciteBstWouldAddEndPuncttrue
\mciteSetBstMidEndSepPunct{\mcitedefaultmidpunct}
{\mcitedefaultendpunct}{\mcitedefaultseppunct}\relax
\EndOfBibitem
\bibitem[Xie and Gao(2007)Xie, and Gao]{X-Pol2007}
Xie,~W.; Gao,~J. Design of a next generation force field: the X-POL potential.
  \emph{J. Chem. Theory Comput.} \textbf{2007}, \emph{3}, 1890--1900\relax
\mciteBstWouldAddEndPuncttrue
\mciteSetBstMidEndSepPunct{\mcitedefaultmidpunct}
{\mcitedefaultendpunct}{\mcitedefaultseppunct}\relax
\EndOfBibitem
\bibitem[Dapprich \latin{et~al.}(1999)Dapprich, Kom{\'a}romi, Byun, Morokuma,
  and Frisch]{ONIOM1}
Dapprich,~S.; Kom{\'a}romi,~I.; Byun,~K.~S.; Morokuma,~K.; Frisch,~M.~J. A new
  ONIOM implementation in Gaussian98. Part I. The calculation of energies,
  gradients, vibrational frequencies and electric field derivatives. \emph{J.
  Mol. Struct. (THEOCHEM)} \textbf{1999}, \emph{461}, 1--21\relax
\mciteBstWouldAddEndPuncttrue
\mciteSetBstMidEndSepPunct{\mcitedefaultmidpunct}
{\mcitedefaultendpunct}{\mcitedefaultseppunct}\relax
\EndOfBibitem
\bibitem[Vreven \latin{et~al.}(2006)Vreven, Byun, Kom{\'a}romi, Dapprich,
  Montgomery~Jr, Morokuma, and Frisch]{ONIOM2}
Vreven,~T.; Byun,~K.~S.; Kom{\'a}romi,~I.; Dapprich,~S.; Montgomery~Jr,~J.~A.;
  Morokuma,~K.; Frisch,~M.~J. Combining quantum mechanics methods with
  molecular mechanics methods in ONIOM. \emph{J. Chem. Theory Comput.}
  \textbf{2006}, \emph{2}, 815--826\relax
\mciteBstWouldAddEndPuncttrue
\mciteSetBstMidEndSepPunct{\mcitedefaultmidpunct}
{\mcitedefaultendpunct}{\mcitedefaultseppunct}\relax
\EndOfBibitem
\bibitem[Guo \latin{et~al.}(2010)Guo, Wu, and Xu]{XO2010}
Guo,~W.; Wu,~A.; Xu,~X. XO: An extended ONIOM method for accurate and efficient
  geometry optimization of large molecules. \emph{Chem. Phys. Lett.}
  \textbf{2010}, \emph{498}, 203--208\relax
\mciteBstWouldAddEndPuncttrue
\mciteSetBstMidEndSepPunct{\mcitedefaultmidpunct}
{\mcitedefaultendpunct}{\mcitedefaultseppunct}\relax
\EndOfBibitem
\bibitem[Wang \latin{et~al.}(2012)Wang, Sosa, Cembran, Truhlar, and
  Gao]{X-Pol2012}
Wang,~Y.; Sosa,~C.~P.; Cembran,~A.; Truhlar,~D.~G.; Gao,~J. Multilevel X-Pol: A
  fragment-based method with mixed quantum mechanical representations of
  different fragments. \emph{J. Phys. Chem. B} \textbf{2012}, \emph{116},
  6781--6788\relax
\mciteBstWouldAddEndPuncttrue
\mciteSetBstMidEndSepPunct{\mcitedefaultmidpunct}
{\mcitedefaultendpunct}{\mcitedefaultseppunct}\relax
\EndOfBibitem
\bibitem[Yang(1991)]{YangDC1991}
Yang,~W. Direct calculation of electron density in density-functional theory.
  \emph{Phys. Rev. Lett.} \textbf{1991}, \emph{66}, 1438\relax
\mciteBstWouldAddEndPuncttrue
\mciteSetBstMidEndSepPunct{\mcitedefaultmidpunct}
{\mcitedefaultendpunct}{\mcitedefaultseppunct}\relax
\EndOfBibitem
\bibitem[Yang and Lee(1995)Yang, and Lee]{YangDC1995}
Yang,~W.; Lee,~T.-S. A density-matrix divide-and-conquer approach for
  electronic structure calculations of large molecules. \emph{J. Chem. Phys.}
  \textbf{1995}, \emph{103}, 5674--5678\relax
\mciteBstWouldAddEndPuncttrue
\mciteSetBstMidEndSepPunct{\mcitedefaultmidpunct}
{\mcitedefaultendpunct}{\mcitedefaultseppunct}\relax
\EndOfBibitem
\bibitem[Akama \latin{et~al.}(2007)Akama, Kobayashi, and Nakai]{NakaiDC2007}
Akama,~T.; Kobayashi,~M.; Nakai,~H. Implementation of divide-and-conquer method
  including Hartree-Fock exchange interaction. \emph{J. Comput. Chem.}
  \textbf{2007}, \emph{28}, 2003--2012\relax
\mciteBstWouldAddEndPuncttrue
\mciteSetBstMidEndSepPunct{\mcitedefaultmidpunct}
{\mcitedefaultendpunct}{\mcitedefaultseppunct}\relax
\EndOfBibitem
\bibitem[He and Merz~Jr(2010)He, and Merz~Jr]{MerzDC2010}
He,~X.; Merz~Jr,~K.~M. Divide and conquer Hartree- Fock calculations on
  proteins. \emph{J. Chem. Theory Comput.} \textbf{2010}, \emph{6},
  405--411\relax
\mciteBstWouldAddEndPuncttrue
\mciteSetBstMidEndSepPunct{\mcitedefaultmidpunct}
{\mcitedefaultendpunct}{\mcitedefaultseppunct}\relax
\EndOfBibitem
\bibitem[Wesolowski and Warshel(1993)Wesolowski, and Warshel]{SubDFT1993}
Wesolowski,~T.~A.; Warshel,~A. Frozen density functional approach for ab initio
  calculations of solvated molecules. \emph{J. Phys. Chem.} \textbf{1993},
  \emph{97}, 8050--8053\relax
\mciteBstWouldAddEndPuncttrue
\mciteSetBstMidEndSepPunct{\mcitedefaultmidpunct}
{\mcitedefaultendpunct}{\mcitedefaultseppunct}\relax
\EndOfBibitem
\bibitem[Assfeld and Rivail(1996)Assfeld, and Rivail]{LSCF1996}
Assfeld,~X.; Rivail,~J.-L. Quantum chemical computations on parts of large
  molecules: the ab initio local self consistent field method. \emph{Chem.
  Phys. Lett.} \textbf{1996}, \emph{263}, 100--106\relax
\mciteBstWouldAddEndPuncttrue
\mciteSetBstMidEndSepPunct{\mcitedefaultmidpunct}
{\mcitedefaultendpunct}{\mcitedefaultseppunct}\relax
\EndOfBibitem
\bibitem[Iannuzzi \latin{et~al.}(2006)Iannuzzi, Kirchner, and
  Hutter]{SubDFT2006}
Iannuzzi,~M.; Kirchner,~B.; Hutter,~J. Density functional embedding for
  molecular systems. \emph{Chem. Phys. Lett.} \textbf{2006}, \emph{421},
  16--20\relax
\mciteBstWouldAddEndPuncttrue
\mciteSetBstMidEndSepPunct{\mcitedefaultmidpunct}
{\mcitedefaultendpunct}{\mcitedefaultseppunct}\relax
\EndOfBibitem
\bibitem[Fux \latin{et~al.}(2010)Fux, Jacob, Neugebauer, Visscher, and
  Reiher]{FDE-pot2010}
Fux,~S.; Jacob,~C.~R.; Neugebauer,~J.; Visscher,~L.; Reiher,~M. Accurate
  frozen-density embedding potentials as a first step towards a subsystem
  description of covalent bonds. \emph{J. Chem. Phys.} \textbf{2010},
  \emph{132}, 164101\relax
\mciteBstWouldAddEndPuncttrue
\mciteSetBstMidEndSepPunct{\mcitedefaultmidpunct}
{\mcitedefaultendpunct}{\mcitedefaultseppunct}\relax
\EndOfBibitem
\bibitem[Chulhai and Jensen(2015)Chulhai, and Jensen]{FDE-EO2015}
Chulhai,~D.~V.; Jensen,~L. Frozen Density Embedding with External Orthogonality
  in Delocalized Covalent Systems. \emph{J. Chem. Theory Comput.}
  \textbf{2015}, \emph{11}, 3080--3088\relax
\mciteBstWouldAddEndPuncttrue
\mciteSetBstMidEndSepPunct{\mcitedefaultmidpunct}
{\mcitedefaultendpunct}{\mcitedefaultseppunct}\relax
\EndOfBibitem
\bibitem[Huang \latin{et~al.}(2011)Huang, Pavone, and
  Carter]{Carter2011quantum}
Huang,~C.; Pavone,~M.; Carter,~E.~A. Quantum mechanical embedding theory based
  on a unique embedding potential. \emph{J. Chem. Phys.} \textbf{2011},
  \emph{134}, 154110\relax
\mciteBstWouldAddEndPuncttrue
\mciteSetBstMidEndSepPunct{\mcitedefaultmidpunct}
{\mcitedefaultendpunct}{\mcitedefaultseppunct}\relax
\EndOfBibitem
\bibitem[Manby \latin{et~al.}(2012)Manby, Stella, Goodpaster, and
  Miller~III]{MFEmbedding2012}
Manby,~F.~R.; Stella,~M.; Goodpaster,~J.~D.; Miller~III,~T.~F. A simple, exact
  density-functional-theory embedding scheme. \emph{J. Chem. Theory Comput.}
  \textbf{2012}, \emph{8}, 2564--2568\relax
\mciteBstWouldAddEndPuncttrue
\mciteSetBstMidEndSepPunct{\mcitedefaultmidpunct}
{\mcitedefaultendpunct}{\mcitedefaultseppunct}\relax
\EndOfBibitem
\bibitem[Knizia and Chan(2012)Knizia, and Chan]{DMET2012}
Knizia,~G.; Chan,~G. K.-L. Density matrix embedding: A simple alternative to
  dynamical mean-field theory. \emph{Phys. Rev. Lett.} \textbf{2012},
  \emph{109}, 186404\relax
\mciteBstWouldAddEndPuncttrue
\mciteSetBstMidEndSepPunct{\mcitedefaultmidpunct}
{\mcitedefaultendpunct}{\mcitedefaultseppunct}\relax
\EndOfBibitem
\bibitem[Knizia and Chan(2013)Knizia, and Chan]{DMET2013}
Knizia,~G.; Chan,~G. K.-L. Density matrix embedding: A strong-coupling quantum
  embedding theory. \emph{J. Chem. Theory Comput.} \textbf{2013}, \emph{9},
  1428--1432\relax
\mciteBstWouldAddEndPuncttrue
\mciteSetBstMidEndSepPunct{\mcitedefaultmidpunct}
{\mcitedefaultendpunct}{\mcitedefaultseppunct}\relax
\EndOfBibitem
\bibitem[Tamukong \latin{et~al.}(2014)Tamukong, Khait, and
  Hoffmann]{HoffmannEmbedding2014}
Tamukong,~P.~K.; Khait,~Y.~G.; Hoffmann,~M.~R. Density differences in embedding
  theory with external orbital orthogonality. \emph{J. Phys. Chem. A}
  \textbf{2014}, \emph{118}, 9182--9200\relax
\mciteBstWouldAddEndPuncttrue
\mciteSetBstMidEndSepPunct{\mcitedefaultmidpunct}
{\mcitedefaultendpunct}{\mcitedefaultseppunct}\relax
\EndOfBibitem
\bibitem[Welborn \latin{et~al.}(2016)Welborn, Tsuchimochi, and
  Van~Voorhis]{Bootstrap2016}
Welborn,~M.; Tsuchimochi,~T.; Van~Voorhis,~T. Bootstrap embedding: An
  internally consistent fragment-based method. \emph{J. Chem. Phys.}
  \textbf{2016}, \emph{145}, 074102\relax
\mciteBstWouldAddEndPuncttrue
\mciteSetBstMidEndSepPunct{\mcitedefaultmidpunct}
{\mcitedefaultendpunct}{\mcitedefaultseppunct}\relax
\EndOfBibitem
\bibitem[H{\'e}gely \latin{et~al.}(2016)H{\'e}gely, Nagy, Ferenczy, and
  K{\'a}llay]{WFTinDFT-LMO}
H{\'e}gely,~B.; Nagy,~P.~R.; Ferenczy,~G.~G.; K{\'a}llay,~M. Exact density
  functional and wave function embedding schemes based on orbital localization.
  \emph{J. Chem. Phys.} \textbf{2016}, \emph{145}, 064107\relax
\mciteBstWouldAddEndPuncttrue
\mciteSetBstMidEndSepPunct{\mcitedefaultmidpunct}
{\mcitedefaultendpunct}{\mcitedefaultseppunct}\relax
\EndOfBibitem
\bibitem[Culpitt \latin{et~al.}(2017)Culpitt, Brorsen, and
  Hammes-Schiffer]{OCBSE2017}
Culpitt,~T.; Brorsen,~K.~R.; Hammes-Schiffer,~S. Communication: Density
  functional theory embedding with the orthogonality constrained basis set
  expansion procedure. \emph{J. Chem. Phys.} \textbf{2017}, \emph{146},
  211101\relax
\mciteBstWouldAddEndPuncttrue
\mciteSetBstMidEndSepPunct{\mcitedefaultmidpunct}
{\mcitedefaultendpunct}{\mcitedefaultseppunct}\relax
\EndOfBibitem
\bibitem[Chen \latin{et~al.}(2019)Chen, Baer, Neuhauser, and
  Rabani]{o-efsDFT2019}
Chen,~M.; Baer,~R.; Neuhauser,~D.; Rabani,~E. Overlapped embedded fragment
  stochastic density functional theory for covalently-bonded materials.
  \emph{J. Chem. Phys.} \textbf{2019}, \emph{150}, 034106\relax
\mciteBstWouldAddEndPuncttrue
\mciteSetBstMidEndSepPunct{\mcitedefaultmidpunct}
{\mcitedefaultendpunct}{\mcitedefaultseppunct}\relax
\EndOfBibitem
\bibitem[Niffenegger \latin{et~al.}(2019)Niffenegger, Oueis, Nafziger, and
  Wasserman]{ChemPotEmb2019}
Niffenegger,~K.; Oueis,~Y.; Nafziger,~J.; Wasserman,~A. Density embedding with
  constrained chemical potential. \emph{Mol. Phys.} \textbf{2019}, \emph{117},
  2188--2194\relax
\mciteBstWouldAddEndPuncttrue
\mciteSetBstMidEndSepPunct{\mcitedefaultmidpunct}
{\mcitedefaultendpunct}{\mcitedefaultseppunct}\relax
\EndOfBibitem
\bibitem[Macetti \latin{et~al.}(2021)Macetti, Wieduwilt, and
  Genoni]{QM-ELMO2021}
Macetti,~G.; Wieduwilt,~E.~K.; Genoni,~A. QM/ELMO: A Multi-Purpose Fully
  Quantum Mechanical Embedding Scheme Based on Extremely Localized Molecular
  Orbitals. \emph{J. Phys. Chem. A} \textbf{2021}, \emph{125}, 2709--2726\relax
\mciteBstWouldAddEndPuncttrue
\mciteSetBstMidEndSepPunct{\mcitedefaultmidpunct}
{\mcitedefaultendpunct}{\mcitedefaultseppunct}\relax
\EndOfBibitem
\bibitem[Stoll \latin{et~al.}(1980)Stoll, Wagenblast, and
  Preu$\beta$]{SCF-MI1980}
Stoll,~H.; Wagenblast,~G.; Preu$\beta$,~H. On the use of local basis sets for
  localized molecular orbitals. \emph{Theor. Chim. Acta} \textbf{1980},
  \emph{57}, 169--178\relax
\mciteBstWouldAddEndPuncttrue
\mciteSetBstMidEndSepPunct{\mcitedefaultmidpunct}
{\mcitedefaultendpunct}{\mcitedefaultseppunct}\relax
\EndOfBibitem
\bibitem[Stoll(1992)]{Incremental1}
Stoll,~H. The correlation energy of crystalline silicon. \emph{Chem. Phys.
  Lett.} \textbf{1992}, \emph{191}, 548--552\relax
\mciteBstWouldAddEndPuncttrue
\mciteSetBstMidEndSepPunct{\mcitedefaultmidpunct}
{\mcitedefaultendpunct}{\mcitedefaultseppunct}\relax
\EndOfBibitem
\bibitem[Friedrich \latin{et~al.}(2007)Friedrich, Hanrath, and
  Dolg]{Incremental2}
Friedrich,~J.; Hanrath,~M.; Dolg,~M. Fully automated implementation of the
  incremental scheme: Application to CCSD energies for hydrocarbons and
  transition metal compounds. \emph{J. Chem. Phys.} \textbf{2007}, \emph{126},
  154110\relax
\mciteBstWouldAddEndPuncttrue
\mciteSetBstMidEndSepPunct{\mcitedefaultmidpunct}
{\mcitedefaultendpunct}{\mcitedefaultseppunct}\relax
\EndOfBibitem
\bibitem[Li \latin{et~al.}(2002)Li, Ma, and Jiang]{CIM2002}
Li,~S.; Ma,~J.; Jiang,~Y. Linear scaling local correlation approach for solving
  the coupled cluster equations of large systems. \emph{J. Comput. Chem.}
  \textbf{2002}, \emph{23}, 237--244\relax
\mciteBstWouldAddEndPuncttrue
\mciteSetBstMidEndSepPunct{\mcitedefaultmidpunct}
{\mcitedefaultendpunct}{\mcitedefaultseppunct}\relax
\EndOfBibitem
\bibitem[Li and Piecuch(2010)Li, and Piecuch]{CIM2010}
Li,~W.; Piecuch,~P. Improved design of orbital domains within the
  cluster-in-molecule local correlation framework: Single-environment
  cluster-in-molecule ansatz and its application to local coupled-cluster
  approach with singles and doubles. \emph{J. Phys. Chem. A} \textbf{2010},
  \emph{114}, 8644--8657\relax
\mciteBstWouldAddEndPuncttrue
\mciteSetBstMidEndSepPunct{\mcitedefaultmidpunct}
{\mcitedefaultendpunct}{\mcitedefaultseppunct}\relax
\EndOfBibitem
\bibitem[Flocke and Bartlett(2004)Flocke, and Bartlett]{NLCC2004}
Flocke,~N.; Bartlett,~R.~J. A natural linear scaling coupled-cluster method.
  \emph{J. Chem. Phys.} \textbf{2004}, \emph{121}, 10935--10944\relax
\mciteBstWouldAddEndPuncttrue
\mciteSetBstMidEndSepPunct{\mcitedefaultmidpunct}
{\mcitedefaultendpunct}{\mcitedefaultseppunct}\relax
\EndOfBibitem
\bibitem[Hughes \latin{et~al.}(2008)Hughes, Flocke, and Bartlett]{NLCC2008}
Hughes,~T.~F.; Flocke,~N.; Bartlett,~R.~J. Natural linear-scaled
  coupled-cluster theory with local transferable triple excitations:
  Applications to peptides. \emph{J. Phys. Chem. A} \textbf{2008}, \emph{112},
  5994--6003\relax
\mciteBstWouldAddEndPuncttrue
\mciteSetBstMidEndSepPunct{\mcitedefaultmidpunct}
{\mcitedefaultendpunct}{\mcitedefaultseppunct}\relax
\EndOfBibitem
\bibitem[Zi{\'o}{\l}kowski \latin{et~al.}(2010)Zi{\'o}{\l}kowski, Jansik,
  Kj{\ae}rgaard, and J{\o}rgensen]{DEC2010}
Zi{\'o}{\l}kowski,~M.; Jansik,~B.; Kj{\ae}rgaard,~T.; J{\o}rgensen,~P. Linear
  scaling coupled cluster method with correlation energy based error control.
  \emph{J. Chem. Phys.} \textbf{2010}, \emph{133}, 014107\relax
\mciteBstWouldAddEndPuncttrue
\mciteSetBstMidEndSepPunct{\mcitedefaultmidpunct}
{\mcitedefaultendpunct}{\mcitedefaultseppunct}\relax
\EndOfBibitem
\bibitem[Jacobson and Herbert(2011)Jacobson, and Herbert]{XPol-SAPT2011}
Jacobson,~L.~D.; Herbert,~J.~M. An efficient, fragment-based electronic
  structure method for molecular systems: Self-consistent polarization with
  perturbative two-body exchange and dispersion. \emph{J. Chem. Phys.}
  \textbf{2011}, \emph{134}, 094118\relax
\mciteBstWouldAddEndPuncttrue
\mciteSetBstMidEndSepPunct{\mcitedefaultmidpunct}
{\mcitedefaultendpunct}{\mcitedefaultseppunct}\relax
\EndOfBibitem
\bibitem[Aoki and Gu(2012)Aoki, and Gu]{ElongationPCCP2012}
Aoki,~Y.; Gu,~F.~L. An elongation method for large systems toward bio-systems.
  \emph{Phys. Chem. Chem. Phys.} \textbf{2012}, \emph{14}, 7640--7668\relax
\mciteBstWouldAddEndPuncttrue
\mciteSetBstMidEndSepPunct{\mcitedefaultmidpunct}
{\mcitedefaultendpunct}{\mcitedefaultseppunct}\relax
\EndOfBibitem
\bibitem[Gadre \latin{et~al.}(1994)Gadre, Shirsat, and Limaye]{CG-MTA1994}
Gadre,~S.~R.; Shirsat,~R.~N.; Limaye,~A.~C. Molecular tailoring approach for
  simulation of electrostatic properties. \emph{J. Phys. Chem.} \textbf{1994},
  \emph{98}, 9165--9169\relax
\mciteBstWouldAddEndPuncttrue
\mciteSetBstMidEndSepPunct{\mcitedefaultmidpunct}
{\mcitedefaultendpunct}{\mcitedefaultseppunct}\relax
\EndOfBibitem
\bibitem[Ganesh \latin{et~al.}(2006)Ganesh, Dongare, Balanarayan, and
  Gadre]{CG-MTA2006}
Ganesh,~V.; Dongare,~R.~K.; Balanarayan,~P.; Gadre,~S.~R. Molecular tailoring
  approach for geometry optimization of large molecules: Energy evaluation and
  parallelization strategies. \emph{J. Chem. Phys.} \textbf{2006}, \emph{125},
  104109\relax
\mciteBstWouldAddEndPuncttrue
\mciteSetBstMidEndSepPunct{\mcitedefaultmidpunct}
{\mcitedefaultendpunct}{\mcitedefaultseppunct}\relax
\EndOfBibitem
\bibitem[Kitaura \latin{et~al.}(1999)Kitaura, Ikeo, Asada, Nakano, and
  Uebayasi]{FMO1999}
Kitaura,~K.; Ikeo,~E.; Asada,~T.; Nakano,~T.; Uebayasi,~M. Fragment molecular
  orbital method: an approximate computational method for large molecules.
  \emph{Chem. Phys. Lett.} \textbf{1999}, \emph{313}, 701--706\relax
\mciteBstWouldAddEndPuncttrue
\mciteSetBstMidEndSepPunct{\mcitedefaultmidpunct}
{\mcitedefaultendpunct}{\mcitedefaultseppunct}\relax
\EndOfBibitem
\bibitem[Huang \latin{et~al.}(2005)Huang, Massa, and Karle]{huang2005kernel}
Huang,~L.; Massa,~L.; Karle,~J. Kernel energy method illustrated with peptides.
  \emph{Int. J. Quantum Chem.} \textbf{2005}, \emph{103}, 808--817\relax
\mciteBstWouldAddEndPuncttrue
\mciteSetBstMidEndSepPunct{\mcitedefaultmidpunct}
{\mcitedefaultendpunct}{\mcitedefaultseppunct}\relax
\EndOfBibitem
\bibitem[Zhang and Zhang(2003)Zhang, and Zhang]{MFCC2003}
Zhang,~D.~W.; Zhang,~J. Z.~H. Molecular fractionation with conjugate caps for
  full quantum mechanical calculation of protein--molecule interaction energy.
  \emph{J. Chem. Phys.} \textbf{2003}, \emph{119}, 3599--3605\relax
\mciteBstWouldAddEndPuncttrue
\mciteSetBstMidEndSepPunct{\mcitedefaultmidpunct}
{\mcitedefaultendpunct}{\mcitedefaultseppunct}\relax
\EndOfBibitem
\bibitem[Xu \latin{et~al.}(2019)Xu, He, Zhu, and Zhang]{MetalMFCC2019}
Xu,~M.; He,~X.; Zhu,~T.; Zhang,~J. Z.~H. A fragment quantum mechanical method
  for metalloproteins. \emph{J. Chem. Theory Comput.} \textbf{2019}, \emph{15},
  1430--1439\relax
\mciteBstWouldAddEndPuncttrue
\mciteSetBstMidEndSepPunct{\mcitedefaultmidpunct}
{\mcitedefaultendpunct}{\mcitedefaultseppunct}\relax
\EndOfBibitem
\bibitem[Deev and Collins(2005)Deev, and Collins]{SMFA2005}
Deev,~V.; Collins,~M.~A. Approximate ab initio energies by systematic molecular
  fragmentation. \emph{J. Chem. Phys.} \textbf{2005}, \emph{122}, 154102\relax
\mciteBstWouldAddEndPuncttrue
\mciteSetBstMidEndSepPunct{\mcitedefaultmidpunct}
{\mcitedefaultendpunct}{\mcitedefaultseppunct}\relax
\EndOfBibitem
\bibitem[Collins(2012)]{SMFA2012}
Collins,~M.~A. Systematic fragmentation of large molecules by annihilation.
  \emph{Phys. Chem. Chem. Phys.} \textbf{2012}, \emph{14}, 7744--7751\relax
\mciteBstWouldAddEndPuncttrue
\mciteSetBstMidEndSepPunct{\mcitedefaultmidpunct}
{\mcitedefaultendpunct}{\mcitedefaultseppunct}\relax
\EndOfBibitem
\bibitem[Jiang \latin{et~al.}(2006)Jiang, Ma, and
  Jiang]{jiang2006electrostatic}
Jiang,~N.; Ma,~J.; Jiang,~Y. Electrostatic field-adapted molecular
  fractionation with conjugated caps for energy calculations of charged
  biomolecules. \emph{J. Chem. Phys.} \textbf{2006}, \emph{124}, 114112\relax
\mciteBstWouldAddEndPuncttrue
\mciteSetBstMidEndSepPunct{\mcitedefaultmidpunct}
{\mcitedefaultendpunct}{\mcitedefaultseppunct}\relax
\EndOfBibitem
\bibitem[Li \latin{et~al.}(2007)Li, Li, and Jiang]{GEBF1}
Li,~W.; Li,~S.; Jiang,~Y. Generalized energy-based fragmentation approach for
  computing the ground-state energies and properties of large molecules.
  \emph{J. Phys. Chem. A} \textbf{2007}, \emph{111}, 2193--2199\relax
\mciteBstWouldAddEndPuncttrue
\mciteSetBstMidEndSepPunct{\mcitedefaultmidpunct}
{\mcitedefaultendpunct}{\mcitedefaultseppunct}\relax
\EndOfBibitem
\bibitem[Hua \latin{et~al.}(2010)Hua, Hua, and Li]{GEBF2}
Hua,~S.; Hua,~W.; Li,~S. An efficient implementation of the generalized
  energy-based fragmentation approach for general large molecules. \emph{J.
  Phys. Chem. A} \textbf{2010}, \emph{114}, 8126--8134\relax
\mciteBstWouldAddEndPuncttrue
\mciteSetBstMidEndSepPunct{\mcitedefaultmidpunct}
{\mcitedefaultendpunct}{\mcitedefaultseppunct}\relax
\EndOfBibitem
\bibitem[Dahlke and Truhlar(2007)Dahlke, and Truhlar]{EE-MBE2007a}
Dahlke,~E.~E.; Truhlar,~D.~G. Electrostatically embedded many-body expansion
  for large systems, with applications to water clusters. \emph{J. Chem. Theory
  Comput.} \textbf{2007}, \emph{3}, 46--53\relax
\mciteBstWouldAddEndPuncttrue
\mciteSetBstMidEndSepPunct{\mcitedefaultmidpunct}
{\mcitedefaultendpunct}{\mcitedefaultseppunct}\relax
\EndOfBibitem
\bibitem[Dahlke and Truhlar(2007)Dahlke, and Truhlar]{EE-MBE2007b}
Dahlke,~E.~E.; Truhlar,~D.~G. Electrostatically Embedded Many-Body Correlation
  Energy, with Applications to the Calculation of Accurate Second-Order
  M{\o}ller- Plesset Perturbation Theory Energies for Large Water Clusters.
  \emph{J. Chem. Theory Comput.} \textbf{2007}, \emph{3}, 1342--1348\relax
\mciteBstWouldAddEndPuncttrue
\mciteSetBstMidEndSepPunct{\mcitedefaultmidpunct}
{\mcitedefaultendpunct}{\mcitedefaultseppunct}\relax
\EndOfBibitem
\bibitem[\v{R}ez\'{a}\v{c} and Salahub(2010)\v{R}ez\'{a}\v{c}, and
  Salahub]{MFBA2010}
\v{R}ez\'{a}\v{c},~J.; Salahub,~D.~R. Multilevel fragment-based approach
  (MFBA): A novel hybrid computational method for the study of large molecules.
  \emph{J. Chem. Theory Comput.} \textbf{2010}, \emph{6}, 91--99\relax
\mciteBstWouldAddEndPuncttrue
\mciteSetBstMidEndSepPunct{\mcitedefaultmidpunct}
{\mcitedefaultendpunct}{\mcitedefaultseppunct}\relax
\EndOfBibitem
\bibitem[Beran and Nanda(2010)Beran, and Nanda]{HMBI2010}
Beran,~G.~J.; Nanda,~K. Predicting organic crystal lattice energies with
  chemical accuracy. \emph{J. Phys. Chem. Lett.} \textbf{2010}, \emph{1},
  3480--3487\relax
\mciteBstWouldAddEndPuncttrue
\mciteSetBstMidEndSepPunct{\mcitedefaultmidpunct}
{\mcitedefaultendpunct}{\mcitedefaultseppunct}\relax
\EndOfBibitem
\bibitem[Mayhall and Raghavachari(2011)Mayhall, and Raghavachari]{MIM2011}
Mayhall,~N.~J.; Raghavachari,~K. Molecules-in-molecules: An extrapolated
  fragment-based approach for accurate calculations on large molecules and
  materials. \emph{J. Chem. Theory Comput.} \textbf{2011}, \emph{7},
  1336--1343\relax
\mciteBstWouldAddEndPuncttrue
\mciteSetBstMidEndSepPunct{\mcitedefaultmidpunct}
{\mcitedefaultendpunct}{\mcitedefaultseppunct}\relax
\EndOfBibitem
\bibitem[Mayhall and Raghavachari(2012)Mayhall, and Raghavachari]{MOBE2012}
Mayhall,~N.~J.; Raghavachari,~K. Many-overlapping-body (MOB) expansion: A
  generalized many body expansion for nondisjoint monomers in molecular
  fragmentation calculations of covalent molecules. \emph{J. Chem. Theory
  Comput.} \textbf{2012}, \emph{8}, 2669--2675\relax
\mciteBstWouldAddEndPuncttrue
\mciteSetBstMidEndSepPunct{\mcitedefaultmidpunct}
{\mcitedefaultendpunct}{\mcitedefaultseppunct}\relax
\EndOfBibitem
\bibitem[Richard and Herbert(2012)Richard, and Herbert]{GenMBE2012}
Richard,~R.~M.; Herbert,~J.~M. A generalized many-body expansion and a unified
  view of fragment-based methods in electronic structure theory. \emph{J. Chem.
  Phys.} \textbf{2012}, \emph{137}, 064113\relax
\mciteBstWouldAddEndPuncttrue
\mciteSetBstMidEndSepPunct{\mcitedefaultmidpunct}
{\mcitedefaultendpunct}{\mcitedefaultseppunct}\relax
\EndOfBibitem
\bibitem[Le \latin{et~al.}(2012)Le, Tan, Ouyang, and Bettens]{le2012combined}
Le,~H.-A.; Tan,~H.-J.; Ouyang,~J.~F.; Bettens,~R.~P. Combined fragmentation
  method: A simple method for fragmentation of large molecules. \emph{J. Chem.
  Theory Comput.} \textbf{2012}, \emph{8}, 469--478\relax
\mciteBstWouldAddEndPuncttrue
\mciteSetBstMidEndSepPunct{\mcitedefaultmidpunct}
{\mcitedefaultendpunct}{\mcitedefaultseppunct}\relax
\EndOfBibitem
\bibitem[Saha and Raghavachari(2014)Saha, and Raghavachari]{DOD2014}
Saha,~A.; Raghavachari,~K. Dimers of dimers (DOD): A new fragment-based method
  applied to large water clusters. \emph{J. Chem. Theory Comput.}
  \textbf{2014}, \emph{10}, 58--67\relax
\mciteBstWouldAddEndPuncttrue
\mciteSetBstMidEndSepPunct{\mcitedefaultmidpunct}
{\mcitedefaultendpunct}{\mcitedefaultseppunct}\relax
\EndOfBibitem
\bibitem[Gordon \latin{et~al.}(2012)Gordon, Fedorov, Pruitt, and
  Slipchenko]{FragChemRev2012}
Gordon,~M.~S.; Fedorov,~D.~G.; Pruitt,~S.~R.; Slipchenko,~L.~V. Fragmentation
  methods: A route to accurate calculations on large systems. \emph{Chem. Rev.}
  \textbf{2012}, \emph{112}, 632--672\relax
\mciteBstWouldAddEndPuncttrue
\mciteSetBstMidEndSepPunct{\mcitedefaultmidpunct}
{\mcitedefaultendpunct}{\mcitedefaultseppunct}\relax
\EndOfBibitem
\bibitem[He \latin{et~al.}(2014)He, Zhu, Wang, Liu, and Zhang]{MFCC-ACR2014}
He,~X.; Zhu,~T.; Wang,~X.; Liu,~J.; Zhang,~J.~Z. Fragment quantum mechanical
  calculation of proteins and its applications. \emph{Acc. Chem. Res.}
  \textbf{2014}, \emph{47}, 2748--2757\relax
\mciteBstWouldAddEndPuncttrue
\mciteSetBstMidEndSepPunct{\mcitedefaultmidpunct}
{\mcitedefaultendpunct}{\mcitedefaultseppunct}\relax
\EndOfBibitem
\bibitem[Li \latin{et~al.}(2014)Li, Li, and Ma]{GEBF-ACR2014}
Li,~S.; Li,~W.; Ma,~J. Generalized energy-based fragmentation approach and its
  applications to macromolecules and molecular aggregates. \emph{Acc. Chem.
  Res.} \textbf{2014}, \emph{47}, 2712--2720\relax
\mciteBstWouldAddEndPuncttrue
\mciteSetBstMidEndSepPunct{\mcitedefaultmidpunct}
{\mcitedefaultendpunct}{\mcitedefaultseppunct}\relax
\EndOfBibitem
\bibitem[Collins \latin{et~al.}(2014)Collins, Cvitkovic, and
  Bettens]{SMFA-ACR2014}
Collins,~M.~A.; Cvitkovic,~M.~W.; Bettens,~R.~P. The combined fragmentation and
  systematic molecular fragmentation methods. \emph{Acc. Chem. Res.}
  \textbf{2014}, \emph{47}, 2776--2785\relax
\mciteBstWouldAddEndPuncttrue
\mciteSetBstMidEndSepPunct{\mcitedefaultmidpunct}
{\mcitedefaultendpunct}{\mcitedefaultseppunct}\relax
\EndOfBibitem
\bibitem[Tanaka \latin{et~al.}(2014)Tanaka, Mochizuki, Komeiji, Okiyama, and
  Fukuzawa]{FMOPCCP2014}
Tanaka,~S.; Mochizuki,~Y.; Komeiji,~Y.; Okiyama,~Y.; Fukuzawa,~K.
  Electron-correlated fragment-molecular-orbital calculations for biomolecular
  and nano systems. \emph{Phys. Chem. Chem. Phys.} \textbf{2014}, \emph{16},
  10310--10344\relax
\mciteBstWouldAddEndPuncttrue
\mciteSetBstMidEndSepPunct{\mcitedefaultmidpunct}
{\mcitedefaultendpunct}{\mcitedefaultseppunct}\relax
\EndOfBibitem
\bibitem[Gadre \latin{et~al.}(2014)Gadre, Yeole, and Sahu]{gadre2014quantum}
Gadre,~S.~R.; Yeole,~S.~D.; Sahu,~N. Quantum chemical investigations on
  molecular clusters. \emph{Chem. Rev.} \textbf{2014}, \emph{114},
  12132--12173\relax
\mciteBstWouldAddEndPuncttrue
\mciteSetBstMidEndSepPunct{\mcitedefaultmidpunct}
{\mcitedefaultendpunct}{\mcitedefaultseppunct}\relax
\EndOfBibitem
\bibitem[Collins and Bettens(2015)Collins, and Bettens]{FragChemRev2015}
Collins,~M.~A.; Bettens,~R.~P. Energy-based molecular fragmentation methods.
  \emph{Chem. Rev.} \textbf{2015}, \emph{115}, 5607--5642\relax
\mciteBstWouldAddEndPuncttrue
\mciteSetBstMidEndSepPunct{\mcitedefaultmidpunct}
{\mcitedefaultendpunct}{\mcitedefaultseppunct}\relax
\EndOfBibitem
\bibitem[Raghavachari and Saha(2015)Raghavachari, and
  Saha]{raghavachari2015accurate}
Raghavachari,~K.; Saha,~A. Accurate composite and fragment-based quantum
  chemical models for large molecules. \emph{Chem. Rev.} \textbf{2015},
  \emph{115}, 5643--5677\relax
\mciteBstWouldAddEndPuncttrue
\mciteSetBstMidEndSepPunct{\mcitedefaultmidpunct}
{\mcitedefaultendpunct}{\mcitedefaultseppunct}\relax
\EndOfBibitem
\bibitem[Gordon(2017)]{gordon2017fragmentation}
Gordon,~M.~S. \emph{Fragmentation: Toward Accurate Calculations on Complex
  Molecular Systems}; John Wiley \& Sons, 2017\relax
\mciteBstWouldAddEndPuncttrue
\mciteSetBstMidEndSepPunct{\mcitedefaultmidpunct}
{\mcitedefaultendpunct}{\mcitedefaultseppunct}\relax
\EndOfBibitem
\bibitem[Saha and Raghavachari(2015)Saha, and Raghavachari]{MIMMOB2015}
Saha,~A.; Raghavachari,~K. Analysis of different fragmentation strategies on a
  variety of large peptides: Implementation of a low level of theory in
  fragment-based methods can be a crucial factor. \emph{J. Chem. Theory
  Comput.} \textbf{2015}, \emph{11}, 2012--2023\relax
\mciteBstWouldAddEndPuncttrue
\mciteSetBstMidEndSepPunct{\mcitedefaultmidpunct}
{\mcitedefaultendpunct}{\mcitedefaultseppunct}\relax
\EndOfBibitem
\bibitem[Kumar and Iyengar(2019)Kumar, and Iyengar]{LC-MFT2019}
Kumar,~A.; Iyengar,~S.~S. Fragment-based electronic structure for potential
  energy surfaces using a superposition of fragmentation topologies. \emph{J.
  Chem. Theory Comput.} \textbf{2019}, \emph{15}, 5769--5786\relax
\mciteBstWouldAddEndPuncttrue
\mciteSetBstMidEndSepPunct{\mcitedefaultmidpunct}
{\mcitedefaultendpunct}{\mcitedefaultseppunct}\relax
\EndOfBibitem
\bibitem[Shang \latin{et~al.}(2010)Shang, Xiang, Li, and
  Yang]{NAOlinearscaling}
Shang,~H.; Xiang,~H.; Li,~Z.; Yang,~J. Linear scaling electronic structure
  calculations with numerical atomic basis set. \emph{Int. Rev. Phys. Chem.}
  \textbf{2010}, \emph{29}, 665--691\relax
\mciteBstWouldAddEndPuncttrue
\mciteSetBstMidEndSepPunct{\mcitedefaultmidpunct}
{\mcitedefaultendpunct}{\mcitedefaultseppunct}\relax
\EndOfBibitem
\bibitem[Goedecker(1999)]{linearscaling1999}
Goedecker,~S. Linear scaling electronic structure methods. \emph{Rev. Mod.
  Phys.} \textbf{1999}, \emph{71}, 1085--1123\relax
\mciteBstWouldAddEndPuncttrue
\mciteSetBstMidEndSepPunct{\mcitedefaultmidpunct}
{\mcitedefaultendpunct}{\mcitedefaultseppunct}\relax
\EndOfBibitem
\bibitem[Bowler and Miyazaki(2012)Bowler, and Miyazaki]{linearscaling2012}
Bowler,~D.~R.; Miyazaki,~T. {\textbackslash}mathcal$\lbrace$O$\rbrace$(N)
  methods in electronic structure calculations. \emph{Reports on Progress in
  Physics} \textbf{2012}, \emph{75}, 036503\relax
\mciteBstWouldAddEndPuncttrue
\mciteSetBstMidEndSepPunct{\mcitedefaultmidpunct}
{\mcitedefaultendpunct}{\mcitedefaultseppunct}\relax
\EndOfBibitem
\bibitem[Kussmann \latin{et~al.}(2013)Kussmann, Beer, and
  Ochsenfeld]{linearscaling2013}
Kussmann,~J.; Beer,~M.; Ochsenfeld,~C. Linear-scaling self-consistent field
  methods for large molecules. \emph{WIREs Comput. Mol. Sci.} \textbf{2013},
  \emph{3}, 614--636\relax
\mciteBstWouldAddEndPuncttrue
\mciteSetBstMidEndSepPunct{\mcitedefaultmidpunct}
{\mcitedefaultendpunct}{\mcitedefaultseppunct}\relax
\EndOfBibitem
\bibitem[Chiba \latin{et~al.}(2007)Chiba, Fedorov, and Kitaura]{FMO-TDDFT}
Chiba,~M.; Fedorov,~D.~G.; Kitaura,~K. Time-dependent density functional theory
  based upon the fragment molecular orbital method. \emph{J. Chem. Phys.}
  \textbf{2007}, \emph{127}, 104108\relax
\mciteBstWouldAddEndPuncttrue
\mciteSetBstMidEndSepPunct{\mcitedefaultmidpunct}
{\mcitedefaultendpunct}{\mcitedefaultseppunct}\relax
\EndOfBibitem
\bibitem[Khait and Hoffmann(2010)Khait, and
  Hoffmann]{HoffmannEmbeddingExcited2010}
Khait,~Y.~G.; Hoffmann,~M.~R. Embedding theory for excited states. \emph{J.
  Chem. Phys.} \textbf{2010}, \emph{133}, 044107\relax
\mciteBstWouldAddEndPuncttrue
\mciteSetBstMidEndSepPunct{\mcitedefaultmidpunct}
{\mcitedefaultendpunct}{\mcitedefaultseppunct}\relax
\EndOfBibitem
\bibitem[Daday \latin{et~al.}(2013)Daday, K\"{o}nig, Valsson, Neugebauer, and
  Filippi]{SS-EP2013}
Daday,~C.; K\"{o}nig,~C.; Valsson,~O.; Neugebauer,~J.; Filippi,~C.
  State-specific embedding potentials for excitation-energy calculations.
  \emph{J. Chem. Theory Comput.} \textbf{2013}, \emph{9}, 2355--2367\relax
\mciteBstWouldAddEndPuncttrue
\mciteSetBstMidEndSepPunct{\mcitedefaultmidpunct}
{\mcitedefaultendpunct}{\mcitedefaultseppunct}\relax
\EndOfBibitem
\bibitem[Hedeg{\aa}rd \latin{et~al.}(2013)Hedeg{\aa}rd, List, Jensen, and
  Kongsted]{PE-MCSCF2013}
Hedeg{\aa}rd,~E.~D.; List,~N.~H.; Jensen,~H. J.~A.; Kongsted,~J. The
  multi-configuration self-consistent field method within a polarizable
  embedded framework. \emph{J. Chem. Phys.} \textbf{2013}, \emph{139},
  044101\relax
\mciteBstWouldAddEndPuncttrue
\mciteSetBstMidEndSepPunct{\mcitedefaultmidpunct}
{\mcitedefaultendpunct}{\mcitedefaultseppunct}\relax
\EndOfBibitem
\bibitem[Yoshikawa \latin{et~al.}(2013)Yoshikawa, Kobayashi, Fujii, and
  Nakai]{DCexcited2013}
Yoshikawa,~T.; Kobayashi,~M.; Fujii,~A.; Nakai,~H. Novel approach to
  excited-state calculations of large molecules based on divide-and-conquer
  method: application to photoactive yellow protein. \emph{J. Phys. Chem. B}
  \textbf{2013}, \emph{117}, 5565--5573\relax
\mciteBstWouldAddEndPuncttrue
\mciteSetBstMidEndSepPunct{\mcitedefaultmidpunct}
{\mcitedefaultendpunct}{\mcitedefaultseppunct}\relax
\EndOfBibitem
\bibitem[Daday \latin{et~al.}(2014)Daday, K{\"o}nig, Neugebauer, and
  Filippi]{WFTinDFT2014}
Daday,~C.; K{\"o}nig,~C.; Neugebauer,~J.; Filippi,~C. Wavefunction in density
  functional theory embedding for excited states: Which wavefunctions, which
  densities? \emph{ChemPhysChem} \textbf{2014}, \emph{15}, 3205--3217\relax
\mciteBstWouldAddEndPuncttrue
\mciteSetBstMidEndSepPunct{\mcitedefaultmidpunct}
{\mcitedefaultendpunct}{\mcitedefaultseppunct}\relax
\EndOfBibitem
\bibitem[Artiukhin \latin{et~al.}(2015)Artiukhin, Jacob, and
  Neugebauer]{subTDDFT2015}
Artiukhin,~D.~G.; Jacob,~C.~R.; Neugebauer,~J. Excitation energies from
  frozen-density embedding with accurate embedding potentials. \emph{J. Chem.
  Phys.} \textbf{2015}, \emph{142}, 234101\relax
\mciteBstWouldAddEndPuncttrue
\mciteSetBstMidEndSepPunct{\mcitedefaultmidpunct}
{\mcitedefaultendpunct}{\mcitedefaultseppunct}\relax
\EndOfBibitem
\bibitem[Gurunathan \latin{et~al.}(2016)Gurunathan, Acharya, Ghosh, Kosenkov,
  Kaliman, Shao, Krylov, and Slipchenko]{EFPexcited2016}
Gurunathan,~P.~K.; Acharya,~A.; Ghosh,~D.; Kosenkov,~D.; Kaliman,~I.; Shao,~Y.;
  Krylov,~A.~I.; Slipchenko,~L.~V. Extension of the effective fragment
  potential method to macromolecules. \emph{J. Phys. Chem. B} \textbf{2016},
  \emph{120}, 6562--6574\relax
\mciteBstWouldAddEndPuncttrue
\mciteSetBstMidEndSepPunct{\mcitedefaultmidpunct}
{\mcitedefaultendpunct}{\mcitedefaultseppunct}\relax
\EndOfBibitem
\bibitem[Li \latin{et~al.}(2016)Li, Li, Lin, and Li]{GEBFexcited}
Li,~W.; Li,~Y.; Lin,~R.; Li,~S. Generalized Energy-Based Fragmentation Approach
  for Localized Excited States of Large Systems. \emph{J. Phys. Chem. A}
  \textbf{2016}, \emph{120}, 9667--9677\relax
\mciteBstWouldAddEndPuncttrue
\mciteSetBstMidEndSepPunct{\mcitedefaultmidpunct}
{\mcitedefaultendpunct}{\mcitedefaultseppunct}\relax
\EndOfBibitem
\bibitem[Liu \latin{et~al.}(2019)Liu, Sun, Glover, and He]{EE-GMF2019}
Liu,~J.; Sun,~H.; Glover,~W.~J.; He,~X. Prediction of excited-state properties
  of oligoacene crystals using fragment-based quantum mechanical method.
  \emph{J. Phys. Chem. A} \textbf{2019}, \emph{123}, 5407--5417\relax
\mciteBstWouldAddEndPuncttrue
\mciteSetBstMidEndSepPunct{\mcitedefaultmidpunct}
{\mcitedefaultendpunct}{\mcitedefaultseppunct}\relax
\EndOfBibitem
\bibitem[Chen \latin{et~al.}(2019)Chen, Fang, and Cui]{MLMBE}
Chen,~W.-K.; Fang,~W.-H.; Cui,~G. A multi-layer energy-based fragment method
  for excited states and nonadiabatic dynamics. \emph{Phys. Chem. Chem. Phys.}
  \textbf{2019}, \emph{21}, 22695--22699\relax
\mciteBstWouldAddEndPuncttrue
\mciteSetBstMidEndSepPunct{\mcitedefaultmidpunct}
{\mcitedefaultendpunct}{\mcitedefaultseppunct}\relax
\EndOfBibitem
\bibitem[Chen \latin{et~al.}(2019)Chen, Fang, and Cui]{ML-MLMBE}
Chen,~W.-K.; Fang,~W.-H.; Cui,~G. Integrating Machine Learning with Multi-Layer
  Energy-Based Fragment Method for Excited States of Large Systems. \emph{J.
  Phys. Chem. Lett.} \textbf{2019}, \relax
\mciteBstWouldAddEndPunctfalse
\mciteSetBstMidEndSepPunct{\mcitedefaultmidpunct}
{}{\mcitedefaultseppunct}\relax
\EndOfBibitem
\bibitem[Macetti and Genoni(2020)Macetti, and Genoni]{QM-ELMO2020}
Macetti,~G.; Genoni,~A. Quantum Mechanics/Extremely Localized Molecular Orbital
  Embedding Strategy for Excited States: Coupling to Time-Dependent Density
  Functional Theory and Equation-of-Motion Coupled Cluster. \emph{J. Chem.
  Theory Comput.} \textbf{2020}, \emph{16}, 7490--7506\relax
\mciteBstWouldAddEndPuncttrue
\mciteSetBstMidEndSepPunct{\mcitedefaultmidpunct}
{\mcitedefaultendpunct}{\mcitedefaultseppunct}\relax
\EndOfBibitem
\bibitem[Al~Hajj \latin{et~al.}(2005)Al~Hajj, Malrieu, and
  Guih{\'e}ry]{REM-PRB2005}
Al~Hajj,~M.; Malrieu,~J.-P.; Guih{\'e}ry,~N. Renormalized excitonic method in
  terms of block excitations: Application to spin lattices. \emph{Phys. Rev. B}
  \textbf{2005}, \emph{72}, 224412\relax
\mciteBstWouldAddEndPuncttrue
\mciteSetBstMidEndSepPunct{\mcitedefaultmidpunct}
{\mcitedefaultendpunct}{\mcitedefaultseppunct}\relax
\EndOfBibitem
\bibitem[Zhang \latin{et~al.}(2012)Zhang, Malrieu, Ma, and Ma]{REM-JCC2012}
Zhang,~H.; Malrieu,~J.-P.; Ma,~H.; Ma,~J. Implementation of renormalized
  excitonic method at ab initio level. \emph{J. Comput. Chem.} \textbf{2012},
  \emph{33}, 34--43\relax
\mciteBstWouldAddEndPuncttrue
\mciteSetBstMidEndSepPunct{\mcitedefaultmidpunct}
{\mcitedefaultendpunct}{\mcitedefaultseppunct}\relax
\EndOfBibitem
\bibitem[Ma \latin{et~al.}(2012)Ma, Liu, and Ma]{REM-JCP2012}
Ma,~Y.; Liu,~Y.; Ma,~H. A new fragment-based approach for calculating
  electronic excitation energies of large systems. \emph{J. Chem. Phys.}
  \textbf{2012}, \emph{136}, 024113\relax
\mciteBstWouldAddEndPuncttrue
\mciteSetBstMidEndSepPunct{\mcitedefaultmidpunct}
{\mcitedefaultendpunct}{\mcitedefaultseppunct}\relax
\EndOfBibitem
\bibitem[Liu and Herbert(2015)Liu, and Herbert]{ALMO-TDDFT2015}
Liu,~J.; Herbert,~J.~M. An efficient and accurate approximation to
  time-dependent density functional theory for systems of weakly coupled
  monomers. \emph{J. Chem. Phys.} \textbf{2015}, \emph{143}, 034106\relax
\mciteBstWouldAddEndPuncttrue
\mciteSetBstMidEndSepPunct{\mcitedefaultmidpunct}
{\mcitedefaultendpunct}{\mcitedefaultseppunct}\relax
\EndOfBibitem
\bibitem[Liu and Herbert(2016)Liu, and Herbert]{ALMO-TDDFT2016}
Liu,~J.; Herbert,~J.~M. Local excitation approximations to time-dependent
  density functional theory for excitation energies in solution. \emph{J. Chem.
  Theory Comput.} \textbf{2016}, \emph{12}, 157--166\relax
\mciteBstWouldAddEndPuncttrue
\mciteSetBstMidEndSepPunct{\mcitedefaultmidpunct}
{\mcitedefaultendpunct}{\mcitedefaultseppunct}\relax
\EndOfBibitem
\bibitem[Closser \latin{et~al.}(2015)Closser, Ge, Mao, Shao, and
  Head-Gordon]{ALMO-CIS2015}
Closser,~K.~D.; Ge,~Q.; Mao,~Y.; Shao,~Y.; Head-Gordon,~M. Superposition of
  fragment excitations for excited states of large clusters with application to
  helium clusters. \emph{J. Chem. Theory Comput.} \textbf{2015}, \emph{11},
  5791--5803\relax
\mciteBstWouldAddEndPuncttrue
\mciteSetBstMidEndSepPunct{\mcitedefaultmidpunct}
{\mcitedefaultendpunct}{\mcitedefaultseppunct}\relax
\EndOfBibitem
\bibitem[Chulhai and Jensen(2016)Chulhai, and Jensen]{FDE-EO-TDDFT2016}
Chulhai,~D.~V.; Jensen,~L. External orthogonality in subsystem time-dependent
  density functional theory. \emph{Phys. Chem. Chem. Phys.} \textbf{2016},
  \emph{18}, 21032--21039\relax
\mciteBstWouldAddEndPuncttrue
\mciteSetBstMidEndSepPunct{\mcitedefaultmidpunct}
{\mcitedefaultendpunct}{\mcitedefaultseppunct}\relax
\EndOfBibitem
\bibitem[Nakai and Yoshikawa(2017)Nakai, and Yoshikawa]{DCexcited2017}
Nakai,~H.; Yoshikawa,~T. Development of an excited-state calculation method for
  large systems using dynamical polarizability: A divide-and-conquer approach
  at the time-dependent density functional level. \emph{J. Chem. Phys.}
  \textbf{2017}, \emph{146}, 124123\relax
\mciteBstWouldAddEndPuncttrue
\mciteSetBstMidEndSepPunct{\mcitedefaultmidpunct}
{\mcitedefaultendpunct}{\mcitedefaultseppunct}\relax
\EndOfBibitem
\bibitem[Ge \latin{et~al.}(2017)Ge, Mao, White, Epifanovsky, Closser, and
  Head-Gordon]{ALMO-CIS+CT2017}
Ge,~Q.; Mao,~Y.; White,~A.~F.; Epifanovsky,~E.; Closser,~K.~D.; Head-Gordon,~M.
  Simulating the absorption spectra of helium clusters (N= 70, 150, 231, 300)
  using a charge transfer correction to superposition of fragment single
  excitations. \emph{J. Chem. Phys.} \textbf{2017}, \emph{146}, 044111\relax
\mciteBstWouldAddEndPuncttrue
\mciteSetBstMidEndSepPunct{\mcitedefaultmidpunct}
{\mcitedefaultendpunct}{\mcitedefaultseppunct}\relax
\EndOfBibitem
\bibitem[Liu \latin{et~al.}(2014)Liu, Zhang, and Liu]{Triad}
Liu,~J.; Zhang,~Y.; Liu,~W. Photoexcitation of Light-Harvesting {C--P--C60}
  Triads: A FLMO-TD-DFT Study. \emph{J. Chem. Theory Comput.} \textbf{2014},
  \emph{10}, 2436--2448\relax
\mciteBstWouldAddEndPuncttrue
\mciteSetBstMidEndSepPunct{\mcitedefaultmidpunct}
{\mcitedefaultendpunct}{\mcitedefaultseppunct}\relax
\EndOfBibitem
\bibitem[Liu and Hoffmann(2014)Liu, and Hoffmann]{SDS}
Liu,~W.; Hoffmann,~M.~R. SDS: the `static-dynamic-static' framework for
  strongly correlated electrons. \emph{Theor. Chem. Acc.} \textbf{2014},
  \emph{133}, 1481\relax
\mciteBstWouldAddEndPuncttrue
\mciteSetBstMidEndSepPunct{\mcitedefaultmidpunct}
{\mcitedefaultendpunct}{\mcitedefaultseppunct}\relax
\EndOfBibitem
\bibitem[Liu and Hoffmann(2016)Liu, and Hoffmann]{iCI}
Liu,~W.; Hoffmann,~M.~R. iCI: Iterative CI toward full CI. \emph{J. Chem.
  Theory Comput.} \textbf{2016}, \emph{12}, 1169--1178, (E)\textbf{12}, 3000
  (2016).\relax
\mciteBstWouldAddEndPunctfalse
\mciteSetBstMidEndSepPunct{\mcitedefaultmidpunct}
{}{\mcitedefaultseppunct}\relax
\EndOfBibitem
\bibitem[Zhang \latin{et~al.}(2020)Zhang, Liu, and Hoffmann]{iCIPT2}
Zhang,~N.; Liu,~W.; Hoffmann,~M.~R. Iterative Configuration Interaction with
  Selection. \emph{J. Chem. Theory Comput.} \textbf{2020}, \relax
\mciteBstWouldAddEndPunctfalse
\mciteSetBstMidEndSepPunct{\mcitedefaultmidpunct}
{}{\mcitedefaultseppunct}\relax
\EndOfBibitem
\bibitem[Zhang \latin{et~al.}(2021)Zhang, Liu, and Hoffmann]{iCIPT2New}
Zhang,~N.; Liu,~W.; Hoffmann,~M.~R. Further Development of iCIPT2 for Strongly
  Correlated Electrons. \emph{J. Chem. Theory Comput.} \textbf{2021},
  \emph{17}, 949--964\relax
\mciteBstWouldAddEndPuncttrue
\mciteSetBstMidEndSepPunct{\mcitedefaultmidpunct}
{\mcitedefaultendpunct}{\mcitedefaultseppunct}\relax
\EndOfBibitem
\bibitem[Wu \latin{et~al.}(2011)Wu, Liu, Zhang, and Li]{FLMO1}
Wu,~F.; Liu,~W.; Zhang,~Y.; Li,~Z. Linear-scaling time-dependent density
  functional theory based on the idea of “from fragments to molecule”.
  \emph{J. Chem. Theory Comput.} \textbf{2011}, \emph{7}, 3643--3660\relax
\mciteBstWouldAddEndPuncttrue
\mciteSetBstMidEndSepPunct{\mcitedefaultmidpunct}
{\mcitedefaultendpunct}{\mcitedefaultseppunct}\relax
\EndOfBibitem
\bibitem[Li \latin{et~al.}(2017)Li, Liu, and Suo]{FLMO3}
Li,~H.; Liu,~W.; Suo,~B. Localization of open-shell molecular orbitals via
  least change from fragments to molecule. \emph{J. Chem. Phys.} \textbf{2017},
  \emph{146}, 104104\relax
\mciteBstWouldAddEndPuncttrue
\mciteSetBstMidEndSepPunct{\mcitedefaultmidpunct}
{\mcitedefaultendpunct}{\mcitedefaultseppunct}\relax
\EndOfBibitem
\bibitem[Liu \latin{et~al.}(1997)Liu, Hong, Dai, Li, and Dolg]{BDF1}
Liu,~W.; Hong,~G.; Dai,~D.; Li,~L.; Dolg,~M. The {Beijing} 4-component density
  functional theory program package ({BDF}) and its application to {EuO},
  {EuS}, {YbO} and {YbS}. \emph{Theor. Chem. Acc.} \textbf{1997}, \emph{96},
  75--83\relax
\mciteBstWouldAddEndPuncttrue
\mciteSetBstMidEndSepPunct{\mcitedefaultmidpunct}
{\mcitedefaultendpunct}{\mcitedefaultseppunct}\relax
\EndOfBibitem
\bibitem[Zhang \latin{et~al.}(2020)Zhang, Suo, Wang, Zhang, Li, Lei, Zou, Gao,
  Peng, Pu, Xiao, Sun, Wang, Ma, Wang, Guo, and Liu]{BDFrev2020}
Zhang,~Y.; Suo,~B.; Wang,~Z.; Zhang,~N.; Li,~Z.; Lei,~Y.; Zou,~W.; Gao,~J.;
  Peng,~D.; Pu,~Z.; Xiao,~Y.; Sun,~Q.; Wang,~F.; Ma,~Y.; Wang,~X.; Guo,~Y.;
  Liu,~W. BDF: A relativistic electronic structure program package. \emph{J.
  Chem. Phys.} \textbf{2020}, \emph{152}, 064113\relax
\mciteBstWouldAddEndPuncttrue
\mciteSetBstMidEndSepPunct{\mcitedefaultmidpunct}
{\mcitedefaultendpunct}{\mcitedefaultseppunct}\relax
\EndOfBibitem
\bibitem[Wang and Gao(2015)Wang, and Gao]{PHO}
Wang,~Y.; Gao,~J. Projected Hybrid Orbitals: A General QM/MM Method. \emph{J.
  Phys. Chem. B} \textbf{2015}, \emph{119}, 1213--1224\relax
\mciteBstWouldAddEndPuncttrue
\mciteSetBstMidEndSepPunct{\mcitedefaultmidpunct}
{\mcitedefaultendpunct}{\mcitedefaultseppunct}\relax
\EndOfBibitem
\bibitem[Boys(1960)]{BoysLMO}
Boys,~S.~F. Construction of some molecular orbitals to be approximately
  invariant for changes from one molecule to another. \emph{Rev. Mod. Phys.}
  \textbf{1960}, \emph{32}, 296--299\relax
\mciteBstWouldAddEndPuncttrue
\mciteSetBstMidEndSepPunct{\mcitedefaultmidpunct}
{\mcitedefaultendpunct}{\mcitedefaultseppunct}\relax
\EndOfBibitem
\bibitem[Boughton and Pulay(1993)Boughton, and Pulay]{PAO}
Boughton,~J.~W.; Pulay,~P. Comparison of the Boys and Pipek-Mezey localizations
  in the local correlation approach and automatic virtual basis selection.
  \emph{J. Comput. Chem.} \textbf{1993}, \emph{14}, 736--740\relax
\mciteBstWouldAddEndPuncttrue
\mciteSetBstMidEndSepPunct{\mcitedefaultmidpunct}
{\mcitedefaultendpunct}{\mcitedefaultseppunct}\relax
\EndOfBibitem
\bibitem[Kutzelnigg and Liu(2005)Kutzelnigg, and Liu]{X2C2005}
Kutzelnigg,~W.; Liu,~W. Quasirelativistic theory equivalent to fully
  relativistic theory. \emph{J. Chem. Phys.} \textbf{2005}, \emph{123},
  241102\relax
\mciteBstWouldAddEndPuncttrue
\mciteSetBstMidEndSepPunct{\mcitedefaultmidpunct}
{\mcitedefaultendpunct}{\mcitedefaultseppunct}\relax
\EndOfBibitem
\bibitem[Liu and Peng(2009)Liu, and Peng]{X2C2009}
Liu,~W.; Peng,~D. Exact two-component Hamiltonians revisited. \emph{J. Chem.
  Phys.} \textbf{2009}, \emph{131}, 031104\relax
\mciteBstWouldAddEndPuncttrue
\mciteSetBstMidEndSepPunct{\mcitedefaultmidpunct}
{\mcitedefaultendpunct}{\mcitedefaultseppunct}\relax
\EndOfBibitem
\bibitem[Liu(2010)]{LiuMP}
Liu,~W. Ideas of relativistic quantum chemistry. \emph{Mol. Phys.}
  \textbf{2010}, \emph{108}, 1679--1706\relax
\mciteBstWouldAddEndPuncttrue
\mciteSetBstMidEndSepPunct{\mcitedefaultmidpunct}
{\mcitedefaultendpunct}{\mcitedefaultseppunct}\relax
\EndOfBibitem
\bibitem[Liu \latin{et~al.}(2003)Liu, Wang, and Li]{BDF2}
Liu,~W.; Wang,~F.; Li,~L. The Beijing density functional (BDF) program package:
  Methodologies and applications. \emph{J. Theor. Comput. Chem.} \textbf{2003},
  \emph{2}, 257--272\relax
\mciteBstWouldAddEndPuncttrue
\mciteSetBstMidEndSepPunct{\mcitedefaultmidpunct}
{\mcitedefaultendpunct}{\mcitedefaultseppunct}\relax
\EndOfBibitem
\bibitem[Liu \latin{et~al.}(2004)Liu, Wang, and Li]{BDF3}
Liu,~W.; Wang,~F.; Li,~L. In \emph{Recent Advances in Relativistic Molecular
  Theory}; Hirao,~K., Ishikawa,~Y., Eds.; World Scientific: Singapore, 2004; pp
  257--282\relax
\mciteBstWouldAddEndPuncttrue
\mciteSetBstMidEndSepPunct{\mcitedefaultmidpunct}
{\mcitedefaultendpunct}{\mcitedefaultseppunct}\relax
\EndOfBibitem
\bibitem[Liu \latin{et~al.}(2004)Liu, Wang, and Li]{BDFECC}
Liu,~W.; Wang,~F.; Li,~L. In \emph{Encyclopedia of Computational Chemistry};
  von Ragu\'e~Schleyer,~P., Allinger,~N.~L., Clark,~T., Gasteiger,~J.,
  Kollman,~P.~A., Schaefer~III,~H.~F., Eds.; Wiley: Chichester, UK, 2004\relax
\mciteBstWouldAddEndPuncttrue
\mciteSetBstMidEndSepPunct{\mcitedefaultmidpunct}
{\mcitedefaultendpunct}{\mcitedefaultseppunct}\relax
\EndOfBibitem
\bibitem[Becke()]{Becke93}
Becke,~A.~D. Density-Functional Thermochemistry. III. The Role of Exact
  Exchange. \relax
\mciteBstWouldAddEndPunctfalse
\mciteSetBstMidEndSepPunct{\mcitedefaultmidpunct}
{}{\mcitedefaultseppunct}\relax
\EndOfBibitem
\bibitem[Stephens \latin{et~al.}(1994)Stephens, Devlin, Chabalowski, and
  Frisch]{B3LYP}
Stephens,~P.~J.; Devlin,~F.~J.; Chabalowski,~C.~F.; Frisch,~M.~J. Ab initio
  calculation of vibrational absorption and circular dichroism spectra using
  density functional force fields. \emph{J. Phys. Chem.} \textbf{1994},
  \emph{98}, 11623--11627\relax
\mciteBstWouldAddEndPuncttrue
\mciteSetBstMidEndSepPunct{\mcitedefaultmidpunct}
{\mcitedefaultendpunct}{\mcitedefaultseppunct}\relax
\EndOfBibitem
\bibitem[Weigend and Ahlrichs({2005})Weigend, and Ahlrichs]{def2}
Weigend,~F.; Ahlrichs,~R. {Balanced basis sets of split valence, triple zeta
  valence and quadruple zeta valence quality for H to Rn: Design and assessment
  of accuracy}. \emph{{Phys. Chem. Chem. Phys.}} \textbf{{2005}}, \emph{{7}},
  {3297--3305}\relax
\mciteBstWouldAddEndPuncttrue
\mciteSetBstMidEndSepPunct{\mcitedefaultmidpunct}
{\mcitedefaultendpunct}{\mcitedefaultseppunct}\relax
\EndOfBibitem
\bibitem[Amara \latin{et~al.}({2000})Amara, Field, Alhambra, and Gao]{GHO}
Amara,~P.; Field,~M.~J.; Alhambra,~C.; Gao,~J.~L. {The generalized hybrid
  orbital method for combined quantum mechanical/molecular mechanical
  calculations: formulation and tests of the analytical derivatives}.
  \emph{{Theor. Chem. Acc.}} \textbf{{2000}}, \emph{{104}}, 336--343\relax
\mciteBstWouldAddEndPuncttrue
\mciteSetBstMidEndSepPunct{\mcitedefaultmidpunct}
{\mcitedefaultendpunct}{\mcitedefaultseppunct}\relax
\EndOfBibitem
\bibitem[Eckard and Exner(2009)Eckard, and Exner]{SPX}
Eckard,~S.; Exner,~T.~E. Improvements in the generalized hybrid orbital method.
  \emph{Int. J. Quantum Chem.} \textbf{2009}, \emph{109}, 1451--1463\relax
\mciteBstWouldAddEndPuncttrue
\mciteSetBstMidEndSepPunct{\mcitedefaultmidpunct}
{\mcitedefaultendpunct}{\mcitedefaultseppunct}\relax
\EndOfBibitem
\bibitem[Jung \latin{et~al.}(2007)Jung, Choi, Sugita, and Ten-no]{TennoGHO}
Jung,~J.; Choi,~C.~H.; Sugita,~Y.; Ten-no,~S. New implementation of a combined
  quantum mechanical and molecular mechanical method using modified generalized
  hybrid orbitals. \emph{J. Chem. Phys.} \textbf{2007}, \emph{127},
  204102\relax
\mciteBstWouldAddEndPuncttrue
\mciteSetBstMidEndSepPunct{\mcitedefaultmidpunct}
{\mcitedefaultendpunct}{\mcitedefaultseppunct}\relax
\EndOfBibitem
\end{mcitethebibliography}

\end{document}